\documentclass[acmsmall, screen, nonacm]{acmart}

\acmJournal{PACMPL}
\acmVolume{1}
\acmNumber{CONF} %
\acmArticle{1}
\acmYear{2018}
\acmMonth{1}
\acmDOI{} %
\startPage{1}

\setcopyright{none}
\usepackage[utf8]{inputenc}
\usepackage{xcolor,colortbl}
\usepackage{tabularx}
\usepackage{listings}
\usepackage{amsmath}
\usepackage{booktabs}
\usepackage{stmaryrd}
\usepackage[inference]{semantic}
\usepackage{algorithm}
\usepackage[noend]{algpseudocode}
\algrenewcommand\alglinenumber[1]{\tiny #1:}
\usepackage[capitalise]{cleveref}
\usepackage{wrapfig}
\usepackage{makecell}
\usepackage{subcaption}
\usepackage{multirow}
\usepackage{siunitx}
\usepackage{xspace}

\usepackage{pifont}
\newcommand{\cmark}{\text{\ding{51}}}
\newcommand{\xmark}{\text{\ding{55}}}
\usepackage{wasysym} %

\usepackage{pgf-umlcd}

\usepackage{rotating}
\definecolor{red}{RGB}{200,0,0}
\definecolor{orange}{RGB}{128,128,0}
\definecolor{orangewheel}{RGB}{1.0, 0.5, 0.0}
\definecolor{green}{RGB}{0,200,0}
\definecolor{blue}{RGB}{0,0,200}
\definecolor{darkgreen}{rgb}{0,0.4,0}
\definecolor{darkgray}{rgb}{.4,.4,.4}
\definecolor{purple}{rgb}{0.6, 0, 0.6}
\usepackage{ifthen}
\newboolean{showcomments}
\setboolean{showcomments}{true} %
\ifthenelse{\boolean{showcomments}}
{\newcommand{\nbnote}[3]{
  \fcolorbox{gray}{yellow}{\bfseries\sffamily\scriptsize#1}
  {\color{#2} \sffamily\small$\blacktriangleright$\textit{#3}$\blacktriangleleft$}
  }
}
{\newcommand{\nbnote}[3]{}
 
}

\newlength\myindent
\algblockdefx[q]{query}{endQuery}
[1]{\textbf{query} #1}
[0]{}
\algblockdefx[u]{update}{endUpdate}
[1]{\textbf{update} #1}
[0]{}
\algblockdefx[c]{compare}{endCompare}
[1]{\textbf{compare} #1}
[0]{}
\algblockdefx[m]{mergee}{endMergee}
[1]{\textbf{merge} #1}
[0]{}
\algblockdefx[s]{source}{endSource}
[1]{\textbf{atSource} #1}
[0]{}
\algblockdefx[d]{down}{endDown}
[1]{\textbf{downstream} (#1)}

\algnewcommand{\SComment}[1]{\Comment{\mbox{{\footnotesize #1}}}}
\algnewcommand{\speckeyword}[1]{\textbf{#1}}
\newcommand{\indentt}{\hskip1.5em}

\makeatletter
\renewcommand{\ALG@name}{Specification}
\ifthenelse{\equal{\ALG@noend}{t}}%
  {\algtext*{endDown}}
  {}%
\ifthenelse{\equal{\ALG@noend}{t}}%
  {\algtext*{endSource}}
  {}%
\ifthenelse{\equal{\ALG@noend}{t}}%
  {\algtext*{endUpdate}}
  {}%
\ifthenelse{\equal{\ALG@noend}{t}}%
  {\algtext*{endQuery}}
  {}%
\ifthenelse{\equal{\ALG@noend}{t}}%
  {\algtext*{endCompare}}
  {}%
\ifthenelse{\equal{\ALG@noend}{t}}%
  {\algtext*{endMergee}}
  {}%
\makeatother

\newcommand{\VeriFx}{VeriFx\xspace}

\newcommand{\ie}{i.e.\xspace}
\newcommand{\eg}{e.g.\xspace}
\newcommand{\cf}{cf.\xspace}

\newcommand{\lub}{\sqcup_v}
\newcommand{\latticeSmaller}{\leq_v}
\newcommand{\reachable}[1]{reachable(#1)}
\newcommand{\compat}[2]{compatible(#1, #2)}
\newcommand{\Statee}{\Sigma}
\newcommand{\srcPre}{enabledSrc}

\newcommand{\srcPrecond}[2]{\srcPre(#1, #2)}

\newcommand{\Op}{\Statee \rightarrow \Statee}
\newcommand{\Boolean}{\mathbb{B}}

\newcommand\rname[1]{\textsc{#1}}
\newcommand{\infrule}[3]{
    \dfrac{
        \begin{array}{@{}c}
        #1
        \end{array}
    }
    {
        \begin{array}{@{}c}
        #2
        \end{array}
    }
    {\ (\rname{#3})}
}

\newcommand{\infer}[3][]{\infrule{#3}{#2}{#1}}

\newcommand{\vfx}[1]{\mathtt{#1}}
\newcommand{\smt}[1]{\mathsf{#1}}

\makeatletter
\newcommand{\dashover}[2][\mathop]{#1{\mathpalette\df@over{{\dashfill}{#2}}}}
\newcommand{\fillover}[2][\mathop]{#1{\mathpalette\df@over{{\solidfill}{#2}}}}
\newcommand{\df@over}[2]{\df@@over#1#2}
\newcommand\df@@over[3]{%
  \vbox{
    \offinterlineskip
    \ialign{##\cr
      #2{#1}\cr
      \noalign{\kern1pt}
      $\m@th#1#3$\cr
    }
  }%
}
\newcommand{\dashfill}[1]{%
  \kern-.5pt
  \xleaders\hbox{\kern.5pt\vrule height.4pt width \dash@width{#1}\kern.5pt}\hfill
  \kern-.5pt
}
\newcommand{\dash@width}[1]{%
  \ifx#1\displaystyle
    2pt
  \else
    \ifx#1\textstyle
      1.5pt
    \else
      \ifx#1\scriptstyle
        1.25pt
      \else
        \ifx#1\scriptscriptstyle
          1pt
        \fi
      \fi
    \fi
  \fi
}
\newcommand{\solidfill}[1]{\leaders\hrule\hfill}
\makeatother

\newcommand{\applyIfEnabled}[2]{ #1 \cdot #2 }

\newcommand{\zeroOrMore}[1]{\overline{#1}}
\newcommand{\oneOrMore}[1]{\dashover{#1}}%

\newcommand{\varName}{\mathit{x}}
\newcommand{\traitLetter}{\mathit{F}}
\newcommand{\traitName}{\mathit{I}}
\newcommand{\objectLetter}{\mathit{J}}
\newcommand{\objectName}{\mathit{O}}
\newcommand{\valDecl}{\mathit{valDecl}}
\newcommand{\methodDecl}{\mathit{methodDecl}}
\newcommand{\methodOrProof}{\mathit{A}}
\newcommand{\valDeclOrMethodDeclOrMethodOrProof}{\mathit{B}}
\newcommand{\var}{\varName}
\newcommand{\expl}[1]{(\textit{#1}) \quad}
\newcommand{\name}{\mathit{n}}
\newcommand{\typeParam}{\mathit{X}}
\newcommand{\typeParams}{\zeroOrMore{\typeParam}}

\newcommand{\classv}[4]{\mathtt{class}\ #1 \, \langle #2 \rangle \,  (#3) \, \{ \, #4  \,\}}
\newcommand{\classvExtends}[5]{\mathtt{class}\ #1 \, \langle #2 \rangle \,  (#3) \; \mathtt{extends} \; #4 \{ \, #5  \,\}}
\newcommand{\trait}[3]{\mathtt{trait}\ #1 \, \langle #2 \rangle \, \{ \, #3 \,\}}
\newcommand{\traitExtends}[4]{\mathtt{trait}\ #1 \, \langle #2 \rangle  \; \mathtt{extends} \; #3  \, \{ \, #4 \,\}}

\newcommand{\adt}[3]{\mathtt{enum}\ #1 \, \langle #2 \rangle \,  \{ \, #3 \,\}}
\newcommand{\adtt}{\mathit{N}}
\newcommand{\prooff}{\mathit{R}}
\newcommand{\proofv}[3]{\mathtt{proof}\ #1 \, \langle #2 \rangle \,  \{ \, #3 \,\}}
\newcommand{\new}[2]{\mathtt{new}\ #1  (\, #2 \,)}
\newcommand{\invoke}[4]{#1 . #2 \;  \langle #3 \rangle \,  (\, #4 \,)}
\newcommand{\patternv}[2]{#1 \ \mathtt{match} \ \{ #2 \}}
\newcommand{\casev}[2]{\mathtt{case} \ #1 \ \mathtt{\Rightarrow} \ #2}
\newcommand{\varvv}[4]{\vfx{val} \ #1 : #2 \ \vfx{=} \ #3 \ \vfx{in} \ #4} %
\newcommand{\ifv}[3]{\vfx{if} \  #1 \ \vfx{then}\ #2  \ \vfx{else}\ #3}
\newcommand{\methodv}[5]{\vfx{def} \ #1 \, \langle #2\rangle \, ( #3 )  :  #4 \ \vfx{=} \ #5 } %
\newcommand{\lambdav}[2]{ (#1) \Rightarrow #2}
\newcommand{\forallv}[2]{\vfx{forall} \, #1 \centerdot #2}
\newcommand{\existsv}[2]{\vfx{exists} \, #1 \centerdot #2}
\newcommand{\impliesv}[2]{#1 \implies #2}
\newcommand{\accessv}[2]{#1 . #2}
\newcommand{\callv}[2]{#1(#2)}
\newcommand{\funTypeV}[2]{#1 \,  \rightarrow \, #2}
\newcommand{\intv}{\mathtt{int}}
\newcommand{\stringv}{\mathtt{string}}
\newcommand{\boolv}{\mathtt{bool}}
\newcommand{\setv}[1]{\vfx{Set \, \langle #1 \rangle}}
\newcommand{\mapv}[2]{\vfx{Map \, \langle #1, #2 \rangle}}
\newcommand{\arithmeticOperationVfx}[2]{#1 \oplus #2}
\newcommand{\booleanOperationVfx}[2]{#1 \otimes #2}

\newcommand{\concat}[2]{\mbox{\texttt{str\_concat}}(#1, #2)}
\newcommand{\concatt}[3]{\mbox{\texttt{str\_concat}}(#1, #2, #3)}

\newcommand{\seq}{\mathbf{ \; ; \;}}

\newcommand{\ifsmt}[3]{\smt{if} (#1, \, #2, \ #3)}
\newcommand{\lambdasmt}[2]{\lambda (#1) . #2} %
\newcommand{\letsmt}[3]{\smt{let} \, #1 = #2 \, \smt{in} \, #3}
\newcommand{\none}[1]{\smt{None} \langle #1 \rangle ()}
\newcommand{\some}[1]{\smt{Some} (#1)}
\newcommand{\sortsmt}[2]{\smt{sort} \, #1 \, #2  }%
\newcommand{\assertsmt}[1]{\smt{assert} \, #1 }
\newcommand{\constsmt}[2]{\smt{const} \, #1 \, #2}%
\newcommand{\funDefsmt}[5]{\smt{fun} \, #1 \langle #2 \rangle ( #3 ) : #4 = #5} %
\newcommand{\adtsmt}[3]{\smt{adt} \, #1 \langle #2 \rangle \{ #3 \} } %
\newcommand{\ctorsmt}[2]{ #1 ( #2 ) } %
\newcommand{\readsmt}[2]{#1 [ \; #2 \; ]}
\newcommand{\writesmt}[3]{#1 [ \, #2 \, ] \! := \! #3}
\newcommand{\readsmtt}[2]{#1 [ #2 ]}
\newcommand{\writesmtt}[3]{#1 [ #2 ] := #3}
\newcommand{\matchsmt}[2]{\smt{match}(#1, \ #2)}
\newcommand{\callsmt}[2]{#1(#2)}
\newcommand{\callWithTypessmt}[3]{#1 \langle #2 \rangle (#3)}
\newcommand{\accesssmt}[2]{#1.#2}
\newcommand{\forallsmt}[2]{\forall (#1) . #2} %
\newcommand{\existssmt}[2]{\exists (#1) . #2} %
\newcommand{\impliessmt}[2]{#1 \implies #2} %
\newcommand{\intsmt}{\smt{int}}
\newcommand{\stringsmt}{\smt{string}}
\newcommand{\boolsmt}{\smt{bool}}
\newcommand{\casesmt}[2]{\smt{case}( #1, \ #2 ) }
\newcommand{\arithmeticOperationSmt}[2]{#1 \oplus #2}
\newcommand{\booleanOperationSmt}[2]{#1 \otimes #2}
\newcommand{\argsmt}[2]{#1 : #2}

\newcommand{\optionType}[1]{\smt{Option} \langle #1 \rangle}
\newcommand{\customType}[2]{#1 \langle #2 \rangle}

\newcommand{\checkproof}[2]{\mathit{prove} ( #1, #2 )}

\newcommand{\constructor}[2]{#1 \, (#2)}
\newcommand{\logicExp}{\mathit{L}}
\newcommand{\expr}{\mathit{e}}
\newcommand{\exprs}{\zeroOrMore{\expr}}

\newcommand{\type}{\mathit{T}}
\newcommand{\method}{\mathit{M}}
\newcommand{\methods}{\zeroOrMore{\method}}

\newcommand{\exprsmt}{\mathit{e}} %
\newcommand{\arrayType}[2]{\smt{Array} \langle #1, #2 \rangle}

\newcommand{\assertion}{\mathit{R}} %

\newcommand{\FunDef}{\mathit{F}} %

\newcommand{\types}{\zeroOrMore{\type}}

\newcommand{\convtype}[1]{\llbracket #1 \rrbracket_t}
\newcommand{\convexp}[1]{\llbracket #1 \rrbracket}
\newcommand{\convptn}[1]{\mathit{pat}\llbracket #1 \rrbracket}
\newcommand{\convdef}[1]{\mathit{def}\llbracket #1 \rrbracket}
\newcommand{\convmethod}[1]{\mathit{method}\llbracket #1 \rrbracket}
\newcommand{\convset}[1]{\convexp{#1}}
\newcommand{\convmap}[1]{\mathit{map}\llbracket #1 \rrbracket}

\newcommand{\className}{\mathit{C}}
\newcommand{\enumName}{\mathit{E}}
\newcommand{\methodName}{\mathit{m}}
\newcommand{\field}{\mathit{v}}

\newcommand{\ctorName}{\mathit{K}}
\newcommand{\proofName}{\mathit{p}}
\newcommand{\typeParamY}{\mathit{Y}}
\newcommand{\typeParamsY}{\zeroOrMore{\typeParamY}}
\newcommand{\param}{\mathit{x}}
\newcommand{\paramY}{\mathit{y}}
\newcommand{\typeP}{\mathit{P}}
\newcommand{\typeQ}{\mathit{Q}}
\newcommand{\patt}{\mathit{r}}
\newcommand{\checksmt}{\smt{check}}

\newcommand{\tenv}{\Gamma}%
\newcommand{\typ}{\type}%
\newcommand{\typeEnv}{\Delta}
\newcommand{\typing}[4][\typeEnv]{#1 ; #2 \vdash #3 : #4}%

\newcommand{\dom}[1]{dom(#1)}
\newcommand{\gFunTypeV}[3]{\langle #1 \rangle \funTypeV{#2}{#3} }
\newcommand{\typeWF}[1]{ \typeEnv \vdash #1 \mbox{ ok}}

\newcommand{\sortName}{\mathit{S}}
\newcommand{\intName}{\mathit{i}}
\newcommand{\funName}{\mathit{f}}
\newcommand{\adtName}{\mathit{A}}

\crefname{algorithm}{Spec.}{Specs.}
\Crefname{algorithm}{Specification}{Specifications}

\newsavebox{\postBox}
\newsavebox{\postProofBox}

\lstdefinelanguage{VeriFx}{
    morekeywords={class, trait, object, extends, val, def, proof, this, forall, exists, if, else, new, private, protected, enum, match, case, true, false, override},
    sensitive=true, %
    morecomment=[l]{//}, %
    morecomment=[s]{/*}{*/}, %
    morestring=[b]" %
}

\definecolor{dkgreen}{rgb}{0,0.6,0}
\definecolor{gray}{rgb}{0.5,0.5,0.5}
\definecolor{mauve}{rgb}{0.58,0,0.82}

\lstdefinestyle{VFxStyle}{
  frame=tb,
  language=VeriFx,
  numbers=left,
  aboveskip=2mm,
  belowskip=2mm,
  showstringspaces=false,
  columns=flexible,
  basicstyle={\small\ttfamily},
  numberstyle=\tiny\color{gray},
  keywordstyle=\color{blue},
  commentstyle=\color{dkgreen},
  stringstyle=\color{mauve},
  frame=single,
  breaklines=true,
  breakatwhitespace=true,
  tabsize=2,
  captionpos=b,
  escapeinside=`',
  mathescape=true
}

\lstdefinestyle{InTextVFxStyle}{
  language=VeriFx,
  numbers=none,
  aboveskip=2mm,
  belowskip=2mm,
  showstringspaces=false,
  columns=flexible,
  basicstyle={\small\ttfamily},
  numberstyle=\tiny\color{gray},
  keywordstyle=\color{blue},
  commentstyle=\color{dkgreen},
  stringstyle=\color{mauve},
  frame=none,
  breaklines=true,
  breakatwhitespace=true,
  tabsize=2,
  captionpos=b,
  escapeinside=`',
  mathescape=true
}

\lstdefinestyle{ScalaStyle}{
  frame=tb,
  language=Scala,
  numbers=left,
  aboveskip=2mm,
  belowskip=2mm,
  showstringspaces=false,
  columns=flexible,
  basicstyle={\small\ttfamily},
  numberstyle=\tiny\color{gray},
  keywordstyle=\color{blue},
  commentstyle=\color{dkgreen},
  stringstyle=\color{mauve},
  frame=single,
  breaklines=true,
  breakatwhitespace=true,
  tabsize=2,
  captionpos=b,
  escapeinside=`',
  mathescape=true
}

\allowdisplaybreaks %

\begin{document}

\title[\VeriFx: Correct Replicated Data Types for the Masses]{VeriFx: Correct Replicated Data Types for the Masses}         %
\author{Kevin De Porre}
\email{kevin.de.porre@vub.be}
\orcid{0000-0001-5469-1001}
\affiliation{%
  \institution{Vrije Universiteit Brussel}
  \streetaddress{Pleinlaan 2}
  \city{Brussels}
  \country{Belgium}
  \postcode{1050}
}

\author{Carla Ferreira}
\affiliation{%
  \institution{NOVA School of Science and Technology}
  \city{Caparica}
  \country{Portugal}}
\email{carla.ferreira@fct.unl.pt}

\author{Elisa Gonzalez Boix}
\email{egonzale@vub.be}
\affiliation{%
  \institution{Vrije Universiteit Brussel}
  \streetaddress{Pleinlaan 2}
  \city{Brussels}
  \country{Belgium}
  \postcode{1050}
}

\begin{abstract}

Distributed systems adopt weak consistency to ensure high availability and low latency, but state convergence is hard to guarantee due to conflicts.
Experts carefully design replicated data types (RDTs) that resemble sequential data types and embed conflict resolution mechanisms that ensure convergence.
Designing RDTs is challenging as their correctness depends on subtleties such as the ordering of concurrent operations.
Currently, researchers manually verify RDTs, either by paper proofs or using proof assistants.
Unfortunately, paper proofs are subject to reasoning flaws and mechanized proofs verify a formalisation instead of a real-world implementation.
Furthermore, writing mechanized proofs is reserved to verification experts and is extremely time consuming.
To simplify the design, implementation, and verification of RDTs, we propose \VeriFx, a high-level programming language with \emph{automated} proof capabilities.
\VeriFx lets programmers implement RDTs atop functional collections and express correctness properties
that are verified automatically.
Verified RDTs can be transpiled to mainstream languages (currently Scala or JavaScript).
\VeriFx also provides libraries for implementing and verifying Conflict-free Replicated Data Types (CRDTs) and Operational Transformation (OT) functions.
These libraries implement the general execution model of those approaches and define their correctness properties.
We use the libraries to implement and %
verify an extensive portfolio of 35 CRDTs %
and reproduce a study on the correctness of OT functions.

\end{abstract}


\maketitle

\section{Introduction}

Replication is essential to modern distributed systems as it enables fast access times and improves the system's overall scalability, availability, and fault tolerance.
When data is replicated across machines,
replicas must be kept consistent to some extent.
When facing network partitions, replicas cannot remain consistent while also accepting reads and writes, a consequence of the CAP theorem~\cite{CAP,CAP12yearsLater,KleppmannCAP}.
Programmers thus face a trade-off between consistency and availability.
Keeping replicas strongly consistent induces high latencies, poor scalability, and reduced availability since updates must be coordinated, \eg using a distributed consensus algorithm.
By relaxing the consistency guarantees, latencies can be reduced and the overall availability improved, but users may observe temporary inconsistencies between replicas. %
Distributed systems increasingly adopt weak consistency models.
However, concurrent operations may lead to conflicts which must be solved in order to guarantee state convergence.
Consider the case of collaborative text editors.
When a user edits a document, the operation is immediately applied locally on the replica and propagated asynchronously to the other replicas.
Since concurrent edits are applied in different orders at different replicas, states can diverge. %

To ensure convergence,~\citet{OT} proposed a technique called Operational Transformation (OT) which modifies incoming operations against previously executed concurrent operations such that the modified operation preserves the intended effect.
Much work focused on designing transformation functions for collaborative text editing~\cite{OT, ImineOT, ResselOT, SunOT, SuleimanOT}, but it has been shown that all of them (even some with mechanized proofs) are wrong~\cite{ImineOT, liOT, osterOT}. 
Since conflict resolution is hard~\cite{CRDTs,DeltaCRDT,jsonCRDT}, 
researchers now focus on designing replicated data types (RDTs) that serve as basic building blocks for the development of highly available distributed systems.
Such RDTs resemble sequential data types (\eg counters, sets, etc.) but include conflict resolution strategies that guarantee convergence in the presence of conflicts. 
Conflict-free Replicated Data Types (CRDTs)~\cite{CRDTs} are a widely adopted family of RDTs that leverage mathematical properties (such as commutative operations) to avoid conflicts by design.  
Many papers~\cite{CRDTs, shapiro2011comprehensive, DeltaCRDT, jsonCRDT, Baquero17, mergeable, CloudTypes, shapiroStateBasedORSet, bieniusa2012optimized} 
propose new or improved RDT designs and include
a formal specification and/or pseudo code of the RDT together with a manual proof of convergence, mostly paper proofs.
Unfortunately, paper proofs are subject to reasoning flaws.

To avoid the pitfalls of paper proofs,~\citet{zellerVerificationCRDTs} and~\citet{IsabelleCRDTs} propose formal frameworks
to verify the correctness of CRDTs using proof assistants.
However, these frameworks use abstract specifications that are disconnected from actual implementations (\eg Akka's CRDT implementations in Scala).
Hence, a particular implementation may be flawed, even though the specification was proven to be correct.
While interactive proofs are more convincing (because the proof logic is machine-checked), they require significant programmer intervention
which is time consuming and reserved to verification experts~\cite{leino-auto-active-verif,continuousReasoning}.
Recent research efforts %
try to automate (part of) the verification process of CRDTs.
\citet{sureshAutomatedVerificationCRDTs} automatically verify CRDTs under different consistency models but require a first-order logic specification of the CRDTs' operations.
\citet{verificationCRDTsLiquidHaskell} leverage an SMT solver to automate part of the verification process but significant parts still need to be proven manually.
We conclude that the development of RDTs is currently reserved to experts in distributed systems and verification.

To simplify the design and implementation of correct RDTs, we propose \VeriFx,
a functional object-oriented programming language with extensive functional collections including tuples, sets, maps, vectors, and lists.
The collections are immutable which is said to be desirable for the implementation of RDTs and their integration in distributed systems~\cite{immutability}.
\VeriFx features a novel proof construct which enables programmers to express correctness properties that are verified automatically. %
For each proof, \VeriFx derives proof obligations and discharges them using SMT solvers.
Verified RDTs can be transpiled to one of the supported target languages (currently Scala or JavaScript). %
We used \VeriFx to develop libraries for the implementation and verification of CRDT and OT data types.
Internally, these libraries use the proof construct to define the necessary correctness properties.
Programmers can also build their own libraries in \VeriFx.

We designed 
\VeriFx to be reminiscent of existing languages (like Scala) and demonstrate that it is possible to derive automated proofs from real-world RDT implementations.
We argue that the ability to implement RDTs and automatically verify them within the \emph{same} language allows programmers to catch mistakes early during the development process. %
To demonstrate the applicability of \VeriFx, we implemented and verified 35 CRDTs, including well-known CRDTs~\cite{shapiro2011comprehensive, Baquero17, shapiroStateBasedORSet, bieniusa2012optimized, kleppmannMap} and new variants.
From these 35 CRDTs, 34 were verified in a matter of seconds and 1 could not be verified due to its recursive nature. %
We also applied \VeriFx to OT and verified \emph{all} transformation functions %
described by~\citet{ImineOT}, and some unpublished designs~\cite{imineRegAndStack}.

In summary, we make the following contributions:
\begin{itemize}
	\item %
      \VeriFx, the first high-level programming language that enables programmers to implement RDTs by composing functional collections, express correctness properties about those RDTs within the same language, and automatically verify those properties. %
	\item We devise \VeriFx libraries that simplify the implementation of CRDT and OT data types and automatically verify the necessary correctness properties.

	\item We give the first fully automated and mechanized proofs for all but one CRDT proposed by~\citet{shapiro2011comprehensive}, all pure op-based CRDTs~\cite{Baquero17}, and many others. %
	\item We reproduce the study of \citet{ImineOT} regarding the verification of OT functions.
\end{itemize}

\section{Motivation}\label{sec:motivation}

To motivate the need for \VeriFx, consider a distributed system in Scala with replicated data on top of Akka's highly-available distributed key-value store\footnote{\url{https://doc.akka.io/docs/akka/current/distributed-data.html}}.
The store provides built-in CRDTs, \eg sets, counters, etc.
However, our system requires a Two-Phase Set (2PSet) CRDT~\cite{shapiro2011comprehensive} that is not provided by Akka.
We thus need to implement it and verify our implementation. %

\begin{wrapfigure}{r}{0.36\textwidth}
  \vspace{-2mm}
  \begin{center}
    \includegraphics[width=0.34\textwidth]{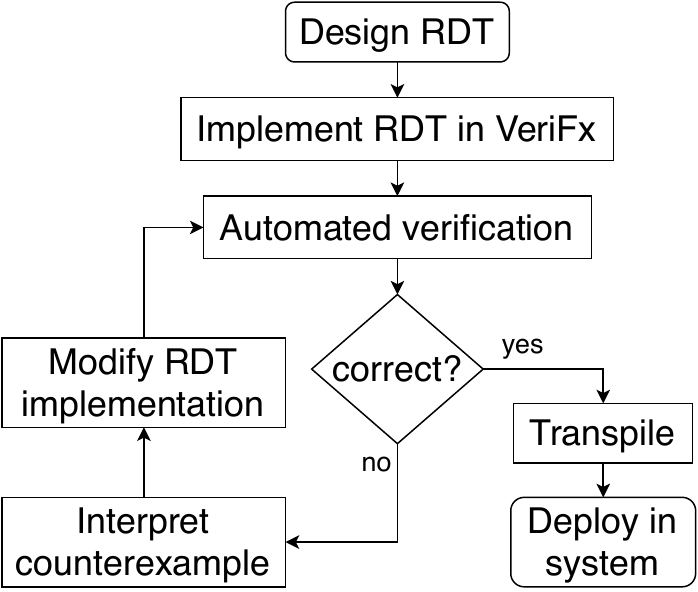}
  \end{center}
  \vspace{-2mm}
  \caption{Workflow for developing RDTs.}
  \label{fig:workflow}
\end{wrapfigure}

Traditionally, software verification requires a complete formalisation of the implementation and its correctness conditions which then need to be proven manually using proof assistants.
The resulting interactive proofs are complex and require much expertise.
For example, \citet{IsabelleCRDTs}'s formalisation and verification of a set CRDT in Isabelle/HOL required the introduction of approximately 20 auxiliary lemmas good for more than 250 LoC in total.
Thus, we cannot reasonably assume that programmers have the time nor the skills to manually verify their implementation using proof assistants~\cite{leino-auto-active-verif,continuousReasoning}.
Alternatively, programmers could resort to \citet{verificationCRDTsLiquidHaskell}'s extension of Liquid Haskell~\cite{liquidHaskell} which automates part of the verification process.
However, non-trivial RDTs still require significant manual proof efforts: $200$+ LoC for a replicated set and $1000$+ LoC for a replicated map~\cite{verificationCRDTsLiquidHaskell}.

In this work, we argue that verification needs to be fully automatic in order to be accessible to non-experts. %
\Cref{fig:workflow} depicts the envisioned workflow for developing RDTs.
Programmers start from a new or existing RDT design and implement it in \VeriFx which will then verify it automatically without the need for a separate formalisation.
If the implementation is not correct, \VeriFx returns a concrete counterexample in which the replicas diverge.
After interpreting the counterexample, the programmer needs to correct the RDT implementation and verify it again.
This iterative process repeats until the implementation is shown correct.
The verified RDT implementation can then be transpiled to a mainstream language (\eg Scala or JavaScript) where it is deployed in the system.

In the remainder of this section we cover each step of the workflow by implementing and verifying an existing 2PSet design in \VeriFx,
transpiling it to Scala, and deploying it on top of Akka.

\subsection{Design and Implementation}\label{sec:motiv-design-and-impl}

\begin{figure*}
\centering
\begin{minipage}[t]{.35\textwidth}
\vspace{-19pt}
\strut\vspace*{-\baselineskip}\newline
\begin{algorithm}[H]
\caption{2PSet CRDT \\taken from~\citet{shapiro2011comprehensive}.}\label{spec:2PSet}
\scriptsize
\begin{algorithmic}[1]
    \State \speckeyword{payload} set $A$, set $R$
 	    \State \indentt \speckeyword{initial} $\varnothing$, $\varnothing$
 	\query{\emph{lookup} (element $e$) : boolean $b$}
 	    \State \speckeyword{let} $b = (e \in A \land e \notin R)$
 	\endQuery
 	\update{\emph{add} (element $e$)}
 	    \State $A := A \cup \{e\}$
 	\endUpdate
 	\update{\emph{remove} (element $e$)}
 	    \State \speckeyword{pre} $\mathit{lookup}(e)$
 	    \State $R := R \cup \{e\}$
 	\endUpdate
 	\compare{($S, T$) : boolean $b$}
        \State \speckeyword{let} $b = (S.A \subseteq T.A \lor S.R \subseteq T.R)$
 	\endCompare
 	\mergee{($S, T$) : payload $U$}
 	  \State \speckeyword{let} $U.A = S.A \cup T.A$
 	  \State \speckeyword{let} $U.R = S.R \cup T.R$
 	\endMergee
 \end{algorithmic}
\end{algorithm}	
\end{minipage}%
\hfill
\begin{minipage}[t]{.58\textwidth}
\vspace{0pt}
\begin{lstlisting}[style=VFxStyle, caption={2PSet implementation in \VeriFx, based on \cref{spec:2PSet}.}, label=lst:2PSet-VFx, basicstyle={\scriptsize\ttfamily}]
class TwoPSet[V](added: Set[V], removed: Set[V]) extends CvRDT[TwoPSet[V]] {
  def lookup(element: V) =
    this.added.contains(element) &&
    !this.removed.contains(element)
  def add(element: V) =
    new TwoPSet(this.added.add(element), this.removed)
  def remove(element: V) =
    new TwoPSet(this.added, this.removed.add(element))
  def compare(that: TwoPSet[V]) =
    this.added.subsetOf(that.added) ||
    this.removed.subsetOf(that.removed)
  def merge(that: TwoPSet[V]) =
    new TwoPSet(this.added.union(that.added), this.removed.union(that.removed))
}
\end{lstlisting}
\end{minipage}
\end{figure*}

\Cref{spec:2PSet} shows the design of the 2PSet CRDT taken from~\citet{shapiro2011comprehensive}.
The 2PSet is a state-based CRDT whose state (the $A$ and $R$ sets) thus forms a join semilattice, \ie a partial order $\latticeSmaller$ with a least upper bound (LUB) $\lub$ for all states.
Elements are added to the 2PSet by adding them to the $A$ set and removed by \emph{adding} them to the $R$ set.
An element is in the 2PSet if it is in $A$ and not in $R$.
Hence, removed elements can never be added again.
Replicas are merged by computing the LUB of their states, which in this case is the union of their respective $A$ and $R$ sets.

The \texttt{compare(S,T)} operation checks if $S \latticeSmaller T$ and is used to define state equivalence: $S \equiv T \iff S \latticeSmaller T \land T \latticeSmaller S$.
Note that state equivalence is defined in terms of $\latticeSmaller$ on the lattice
so that replicas may be considered equivalent even though they are not identical.
This is relevant for CRDTs that keep additional information.
For example, CRDTs often use a lamport clock to generate globally unique IDs.
This lamport clock is different at every replica and is not part of the lattice even though it is part of the state.

\Cref{lst:2PSet-VFx} shows the implementation of the 2PSet CRDT in \VeriFx, which is a straightforward translation of the specification.
The \texttt{TwoPSet} class is polymorphic in the type of values it stores.
It defines the \texttt{added} and \texttt{removed} fields which correspond to the $A$ and $R$ sets respectively. %
The \texttt{add} and \texttt{remove} methods return an updated copy of the state.
The class extends the \texttt{CvRDT} trait\footnote{VeriFx traits can declare abstract methods and fields, and provide default implementations for methods.} that is provided by \VeriFx's CRDT library for building state-based CRDTs (explained later in \cref{sec:crdt-lib}). %
This trait requires the class to implement the \texttt{compare} and \texttt{merge} methods. %

\subsection{Verification}\label{sec:motivation-verif}

We now verify our 2PSet implementation in \VeriFx.
State-based CRDTs guarantee convergence iff the merge function is idempotent, commutative, and associative~\cite{CRDTs}.
\VeriFx's CRDT library includes several \texttt{CvRDTProof} traits which encode these correctness conditions (explained later in \cref{sec:crdt-lib}). %
To verify our \texttt{TwoPSet}, we define a \texttt{TwoPSetProof} object that extends the \texttt{CvRDTProof1} trait and passes the type constructor of the CRDT we want to verify (\ie \texttt{TwoPSet}) as a type argument to the trait: %
\begin{lstlisting}[style=InTextVFxStyle]
object TwoPSetProof extends CvRDTProof1[TwoPSet]
\end{lstlisting}
The \texttt{TwoPSetProof} object inherits an automated correctness proof for the polymorphic \texttt{TwoPSet} CRDT.
When executing this object, \VeriFx will automatically try to verify this proof.
In this case, \VeriFx proves that the \texttt{TwoPSet} guarantees convergence (independent of the type of elements it holds), 
according to the notion of state equivalence that is derived from \texttt{compare}.
However, \VeriFx raises a warning that this notion of equivalence does not correspond to structural equality.
As explained before, this may be normal in some CRDT designs but it requires further investigation.

\VeriFx provides a counterexample consisting of two states 
$S = \texttt{TwoPSet(\{x\},\{\})}$ and $T = \texttt{TwoPSet(\{x\},\{x\})}$ which are considered equivalent $S \equiv T$ but are not identical $S \neq T$.
These two states should indeed not be considered equivalent since $x \in S$ but $x \notin T$ according to \texttt{lookup}.
Looking back at \cref{spec:2PSet}, we notice that \texttt{compare} defines replica $S$ to be smaller or equal to replica $T$ iff $S.A \subseteq T.A$ \emph{or} $S.R \subseteq T.R$.
Since $S.A = T.A$ it follows that $S \latticeSmaller T \land T \latticeSmaller S$ and thus they are considered equal ($S \equiv T$) without even considering the removed elements (i.e. the $R$ sets).
Based on this counterexample, we correct \texttt{compare} such that it considers both the $A$ sets \emph{and} the $R$ sets:
\begin{lstlisting}[style=InTextVFxStyle]
def compare(that: TwoPSet[V]) =
  this.added.subsetOf(that.added) && this.removed.subsetOf(that.removed)
\end{lstlisting}
We verify the implementation again to check that it still guarantees convergence according to this modified definition of equivalence.
\VeriFx automatically proves that the modified implementation is correct and the warning about equivalence is now gone (meaning that the definition of equality that is derived from compare corresponds to structural equality, \ie $s_1 \equiv s_2 \iff s_1 = s_2$).

We completed the verification of the 2PSet CRDT in \VeriFx without providing any verification-specific code.
This example showcases the importance of automated verification as it detected an error in the specification that would have percolated to the implementation.

\subsection{Deployment}

The final step in our workflow consists of automatically transpiling the implementation from \VeriFx to Scala and integrating it in our distributed system which uses Akka's distributed key-value store.

\begin{figure*}
\centering
\begin{minipage}[t]{.52\textwidth}
\begin{lstlisting}[style=ScalaStyle, caption={Transpiled 2PSet in Scala.}, label=lst:2PSet-transpiled, basicstyle={\scriptsize\ttfamily}]
case class TwoPSet[V](added: Set[V], removed: Set[V]) extends CvRDT[TwoPSet[V]] { // CvRDT trait provided by our CRDT library is also compiled to Scala
  def lookup(element: V) = this.added.contains(element) &&
                              !this.removed.contains(element)
  def add(element: V): TwoPSet[V] =
    TwoPSet[V](this.added + element, this.removed)
  def remove(element: V): TwoPSet[V] =
    TwoPSet[V](this.added, this.removed + element)
  def compare(that: TwoPSet[V]): Boolean =
    this.added.subsetOf(that.added) && 
    this.removed.subsetOf(that.removed)
  def merge(that: TwoPSet[V]): TwoPSet[V] =
    TwoPSet[V](this.added.union(that.added), 
                this.removed.union(that.removed)) }
\end{lstlisting}
\end{minipage}
\hfill
\begin{minipage}[t]{.42\textwidth}
\begin{lstlisting}[style=ScalaStyle, caption={Modified 2PSet implementation for integration with Akka's distributed key-value store.}, label=lst:2PSet-akka, basicstyle={\scriptsize\ttfamily}]
@SerialVersionUID(1L)`\label{ln:2pset-serial-version}'
case class TwoPSet[V](
    added: Set[V], removed: Set[V]) extends CvRDT[TwoPSet[V]] with ReplicatedData with Serializable {`\label{ln:2pset-extends-traits}'
  type T = TwoPSet[V]`\label{ln:2PSet-type-member}'
  // The remainder of the implementation is unchanged
}
\end{lstlisting}
\end{minipage}
\vspace{-6mm}
\end{figure*}

\Cref{lst:2PSet-transpiled} shows the transpiled implementation of the 2PSet in Scala.
To store the RDT in Akka's distributed key-value store, this implementation requires two modifications which are shown in \cref{lst:2PSet-akka}.
First, the RDT must extend Akka's \texttt{ReplicatedData} trait (\cref{ln:2pset-extends-traits}) which requires at least the definition of a type member $T$ corresponding to the actual type of the CRDT (\cref{ln:2PSet-type-member}) and a \texttt{merge} method for CRDTs of that type (which we already have).
Second, the RDT must be serializable.
For simplicity, we use Java's built-in serializer\footnote{In production it would be safer and more efficient to implement a custom serializer, \eg using Protobuf.}.
Hence, it suffices to extend the \texttt{Serializable} trait (\cref{ln:2pset-extends-traits}) and to annotate the class with a serial version (\cref{ln:2pset-serial-version}).
After applying these modifications, our verified \texttt{TwoPSet} can be stored in Akka's distributed key-value store and will automatically be replicated across the cluster and be kept eventually consistent.

\section{The \VeriFx Language}\label{sec:lang-def}

 The goal of this work is to build a familiar high-level programming language that is suited to implement RDTs and \emph{automatically} verify them.
The main challenge consists of 
efficiently encoding every feature of the language without breaking automatic verification.
The result of this exercise is \VeriFx, a functional object-oriented programming language 
with Scala-like syntax and a type system that resembles Featherweight Generic Java~\cite{fjToplas}.
\VeriFx features a novel proof construct to express correctness properties about programs.
For every proof construct a proof obligation is derived that is discharged automatically by an SMT solver (\cf \cref{sec:auto-verif}). 

\VeriFx advocates for the object-oriented programming paradigm as it is widespread across programmers and fits the conceptual representation of replicated data as ``shared'' objects.
The functional aspect of the language, in particular its immutable collections, make the language suitable for implementing and integrating RDTs in distributed systems, as argued by~\citet{immutability}.

The remainder of this section is organised in three parts.
First, we give an overview of \VeriFx's architecture.
Second, we define its syntax.
Third, we describe its functional collections.
\VeriFx's type system is described in \cref{app:type-system} as part of the additional material.

\subsection{Overall Architecture}

\begin{figure*}%
	\centering
	\includegraphics[width=0.9\textwidth]{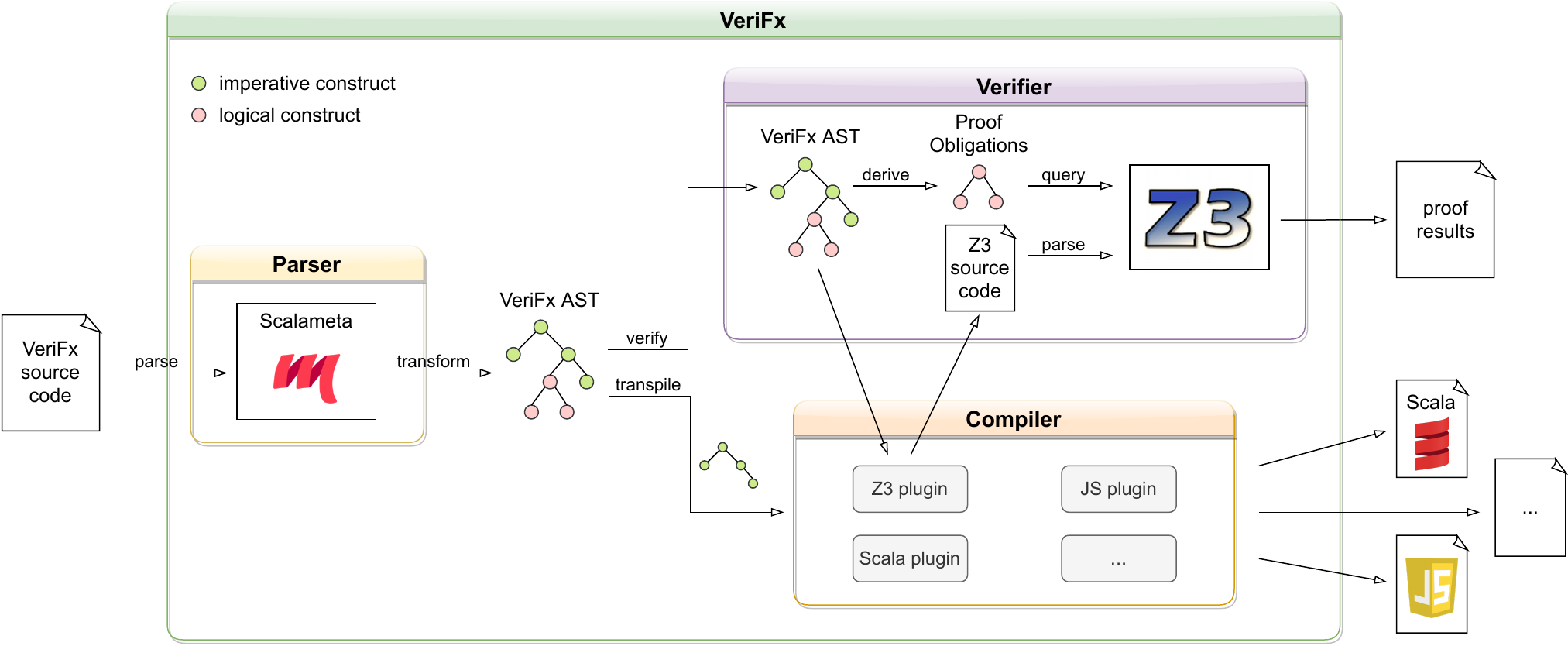}
	\caption{\VeriFx's plugin architecture.}
	\label{fig:vfx-architecture}
	\vspace{-1mm}
\end{figure*}

\Cref{fig:vfx-architecture} provides an overview of \VeriFx's architecture.
Source code is parsed into an Abstract Syntax Tree (AST) representing the program.
Interestingly, every piece of \VeriFx code is valid Scala syntax (but not necessarily semantically correct).
This enables \VeriFx to use Scala Meta\footnote{\url{https://scalameta.org/}} to parse the source code into an AST representing the Scala program, which is then transformed to represent the \VeriFx program.

The AST representing a \VeriFx program can be verified or transpiled to other languages.
Transpilation is done by the compiler which features plugins. %
Support for new languages can be added by implementing a \emph{compiler plugin} for them.
These plugins dictate the compilation of the AST to the target language.
Currently, \VeriFx comes with compiler plugins for Scala, JavaScript, and Z3~\cite{z3} (a state-of-the-art SMT solver).

To verify \VeriFx programs, the verifier derives the necessary proof obligations from the AST.
It then compiles the program to Z3 and automatically discharges the proof obligations.
For every proof, the outcome (accepted, rejected, or unknown) is signaled to the user.
Accepted means that the property holds, rejected means that a counterexample was found for which the property does not hold, and unknown means that the property could not be verified within a certain time frame (which is configurable).
Support for other SMT solvers can also be added by implementing a compiler plugin for them.

\subsection{Syntax}

\Cref{fig:verifx-constructs} defines the syntax of \VeriFx.
The metavariable $\className$ ranges over class names;
$\objectName$ ranges over object names;
$\traitName$ ranges over trait names;
$\enumName$ ranges over enumeration names;
$\ctorName$ ranges over constructor names of enumerations;
$\type$, $\typeP$ and $\typeQ$ range over types;
$\typeParam$ and $\typeParamY$ range over type variables;
$\field$ ranges over field names;
$\param$ and $\paramY$ range over parameter and variable names;
$\methodName$ ranges over method names;
$\proofName$ ranges over proof names;
and $\expr$ ranges over expressions.

\begin{figure}
  \centering
  $$
  \small{
  \begin{array}{@{}r@{\,}c@{\,}l   @{}r@{\,}c@{\,}l}
  L \, & ::= \, & \classv{\className}{\typeParams}{\zeroOrMore{\field} : \zeroOrMore{\type}}{\methods} &
  \qquad
  \method & ::=& \methodv{\methodName}{\zeroOrMore{\typeParam}}{\zeroOrMore{\param} : \zeroOrMore{\type}}{\type}{\expr}  \\   
  & \mid & \classvExtends{\className}{\typeParams}{\zeroOrMore{\field} : \zeroOrMore{\type}}{\traitName \, \langle \, \zeroOrMore{P} \, \rangle }{\methods} &
  \qquad
   \type & ::=  & \intv \, \mid \, \stringv \, \mid \,  \boolv \, \mid \, %
	  \className \, \langle \, \zeroOrMore{T} \, \rangle \, \mid \,
	  \traitName \, \langle \, \zeroOrMore{T} \, \rangle \\
  \objectLetter & ::= & \mathtt{object} \, \objectName \, \{ \, \zeroOrMore{\methodOrProof} \, \} &
  \qquad 
  	         & \mid & \enumName \, \langle \, \zeroOrMore{T} \, \rangle \, \mid \,
	          \funTypeV{\oneOrMore{T}}{T} \\
    & \mid & \mathtt{object} \, \objectName \; \mathtt{extends} \; \traitName \, \langle \, \zeroOrMore{T} \, \rangle \, \{ \, \zeroOrMore{\methodOrProof} \, \} &
    \qquad 
    \expr & ::= & \vfx{num} \, \mid \,  \vfx{str} \, \mid \, \vfx{true} \, \mid \, \vfx{false} %
    \\
  \traitLetter \, & ::= \, & \trait{\traitName}{\zeroOrMore{\typeParam} <: \zeroOrMore{\type}}{\zeroOrMore{\valDeclOrMethodDeclOrMethodOrProof}} &
  \qquad
  & \mid & \arithmeticOperationVfx{\expr}{\expr} \, \mid \,  \booleanOperationVfx{\expr}{\expr} \, \mid \, \, !e \, \mid \, \varName 
  	\, \mid \, \expr . \field \, \mid \, \invoke{\expr}{\methodName}{\zeroOrMore{\type}}{\exprs} 
  \\
     & \mid & \traitExtends{\traitName}{\zeroOrMore{\typeParam} <: \zeroOrMore{\type}}{ \traitName \, \langle \, \zeroOrMore{P} \, \rangle }{\zeroOrMore{\valDeclOrMethodDeclOrMethodOrProof}} &
     \qquad
     & \mid & \varvv{\varName}{\type}{\expr}{\expr} \, \mid \, \ifv{\expr}{\expr}{\expr} 
  \\
  \adtt & ::=& \adt{\enumName}{\zeroOrMore{\typeParam}}{\oneOrMore{\constructor{\ctorName}{\zeroOrMore{\field} : \zeroOrMore{\type}}}} &
  \qquad
  & \mid & \lambdav{\oneOrMore{\param} : \oneOrMore{\type}} {\expr} \, \mid \, \callv{\expr}{\zeroOrMore{\expr}} 
 \\ 
  \methodOrProof & ::= & \method \mid \prooff &
  \qquad
  & \mid & \new{\customType{\className}{\zeroOrMore{\type}}}{\exprs} \, \mid \, \new{\customType{\ctorName}{\zeroOrMore{\type}}}{\exprs} 
  \\
  \valDeclOrMethodDeclOrMethodOrProof & ::= & \valDecl \mid \methodDecl \mid \method \mid \prooff &
  \qquad
  & \mid & \patternv{\expr}{\oneOrMore{\casev{\patt}{\expr}}}
  \\
  \prooff & ::=& \proofv{\proofName}{\zeroOrMore{\typeParam}}{\expr} &
  \qquad  
  & \mid & \forallv {(\oneOrMore{\param} : \oneOrMore{\type})} {\expr} \, \mid \, \existsv {(\oneOrMore{\param} : \oneOrMore{\type})} {\expr} 
  \\ 
  \valDecl & ::= & \mathtt{val} \, \varName : \type &
  \qquad
  & \mid &\impliesv {\expr} {\expr} 
  \\
  \methodDecl & ::= & \mathtt{def} \, \methodName \, \langle \typeParams \rangle (\zeroOrMore{\param} : \zeroOrMore{\type}) : \type &
  \qquad
  \patt & ::= & \callv{\ctorName}{\zeroOrMore{\param}} \, \mid \, \param \, \mid \, \textbf{\_}
  \end{array}}
  $$
  \vspace{-4mm}
  \caption{\VeriFx syntax.}
  \label{fig:verifx-constructs}
  \vspace{-3mm}
\end{figure}

\VeriFx programs consist of one or more statements which can be the definition of an object $\objectName$, a class $\customType{\className}{\typeParams}$, a trait $\customType{\traitName}{\typeParams}$, or an enumeration $\customType{\enumName}{\typeParams}$. %
Objects, classes, enumerations, and traits can be polymorphic and inherit from a single trait (except enumerations).
Objects define zero or more methods and proofs. %
Classes contain zero or more\footnote{An overline, \eg $\zeroOrMore{\typeParam}$, denotes zero or more. A dashed overline, \eg $\oneOrMore{\typeParam}$, denotes one or more.} fields and (polymorphic) methods. %
The body of a method must contain a well-typed expression $\expr$.
Traits can declare values and methods that need to be provided by concrete classes extending the trait,
and define (polymorphic) methods and proofs.
Traits can express upper type bounds on their type parameters to restrict the possible extensions.
Enumerations (enums for short) define one or more constructors, each of which contains zero or more fields.
Programmers can deconstruct enums by pattern matching on them.

Unique to \VeriFx is its proof construct which has a name and whose body must be a well-typed boolean expression.
The body expresses a property that must be verified.
A proof is accepted if its body always evaluates to true, otherwise it is rejected; when rejected, \VeriFx provides a concrete counterexample for which the property does not hold. %
Proofs can be polymorphic, that means they prove a property for all possible type instantiations of their type parameters.
Polymorphic proofs are useful to prove that a polymorphic RDT converges independent of the type of values it contains.

\VeriFx supports a variety of expressions, including
literal values,
arithmetic $\arithmeticOperationVfx{}{}$ and boolean operations $\booleanOperationVfx{}{}$,
boolean negation $!\expr$,
field accesses $\expr . \field$ and method calls $\invoke{\expr}{\methodName}{\zeroOrMore{\type}}{\exprs}$,
variable definitions,
if tests,
anonymous functions and function calls,
class and enum instantiations,
pattern matching,
quantified formulas,
and logical implication.
Functions are \emph{first-class} and take at least one argument because nullary functions are constants.
\VeriFx supports single inheritance from traits to foster code re-use but imposes some limitations.
For example, the arguments of a class method need to be concrete (\ie can not be of a trait type) because proofs about these methods require reasoning about all subtypes but these may not necessarily be known at compile time.
In contrast, enumerations are supported because their constructors are fixed and known at compile time.

\subsection{Functional Collections}\label{sec:collections-overview}

\begin{figure*}
\begin{small}
\begin{tikzpicture}[scale=0.7] %
  \begin{class}[text width=3cm]{Tuple<A, B>}{-5,0}
    \attribute{+ fst : A}
    \attribute{+ snd : B}
  \end{class}
  
  \begin{class}[text width=4.9cm]{Set<V>}{-5,-2.4}
    \operation{+ add(e: V) : Set<V>}
    \operation{+ remove(e: V) : Set<V>}
    \operation{+ contains(e: V) : bool}
    \operation{+ isEmpty() : bool}
    \operation{+ nonEmpty() : bool}
    \operation{+ union(s: Set<V>) : Set<V>}
    \operation{+ diff(s: Set<V>) : Set<V>}
    \operation{+ intersect(s: Set<V>) : Set<V>}
    \operation{+ subsetOf(that: Set[V]) : bool}
    \operation{+ map<W>(f: V => W) : Set<W>}
    \operation{+ filter(p: V => bool) : Set<V>}
    \operation{+ forall(p: V => bool) : bool}
    \operation{+ exists(p: V => bool) : bool}
  \end{class}
  
  \begin{class}[text width=8cm]{Map<K, V>}{5.05,-0.68}
    \operation{+ add(k: K, v: V) : Map<K, V>}
    \operation{+ remove(k: K) : Map<K, V>}
    \operation{+ contains(k: K) : bool}
    \operation{+ get(k: K) : V}
    \operation{+ getOrElse(k: K, default: V) : V}
    \operation{+ keys() : Set<K>}
    \operation{+ values() : Set<V>}
    \operation{+ bijective() : bool}
    \operation{+ map<W>(f: (K, V) => W) : Map<K, W>}
    \operation{+ mapValues<W>(f: V => W) : Map<K, W>}
    \operation{+ filter(p: (K, V) => bool) : Map<K, V>}
    \operation{+ zip<W>(m: Map<K, W>) : Map<K, Tuple<V, W>{}>}
    \operation{+ combine(m: Map<K, V>, f: (V, V) => V) : Map<K, V>}
    \operation{+ forall(p: (K, V) => bool) : bool}
    \operation{+ exists(p: (K, V) => bool) : bool}
    \operation{+ toSet() : Set<Tuple<K, V>>}
  \end{class}
  
  \begin{class}[text width=7cm]{Vector<V>}{-3.5, -10.9} %
    \attribute{+ size : Int}
    \operation{+ get(idx: Int) : V}
    \operation{+ write(idx: Int, value: V) : Vector<V>}
    \operation{+ append(value: V) : Vector<V>}
    \operation{+ map<W>(f: V => W) : Vector<W>}
    \operation{+ zip<W>(v: Vector<W>): Vector<Tuple<V,W>{}>}
    \operation{+ forall(p: V => bool) : bool}
    \operation{+ exists(p: V => bool) : bool}
  \end{class}
  
  \begin{class}[text width=6.1cm]{List<V>}{6.4, -10.9} %
    \attribute{+ size : Int}
    \operation{+ get(idx: Int) : V}
    \operation{+ insert(idx: Int, value: V) : List<V>}
    \operation{+ delete(idx: Int) : List<V>}
    \operation{+ map<W>(f: V => W) : List<W>}
    \operation{+ zip<W>(l: List<W>): List<Tuple<V,W>{}>}
    \operation{+ forall(p: V => bool) : bool}
    \operation{+ exists(p: V => bool) : bool}
  \end{class}
\end{tikzpicture}
\end{small}
\caption{An overview of \VeriFx's built-in functional collections.}
\label{fig:verifx-collections}
\vspace{-0.5mm}
\end{figure*}

\VeriFx features built-in collections for tuples, sets, maps, vectors, and lists.
Remarkably, these collections are completely verifiable and can be arbitrarily composed to build custom RDTs. 
All collections are immutable, ``mutators'' thus return an updated copy of the object.
\Cref{fig:verifx-collections} provides an overview of the interface exposed by these collections, which is heavily inspired by functional programming.
A tuple groups two elements which can be accessed using the \texttt{fst} and \texttt{snd} fields.

\paragraph{Sets.}
Support the typical set operations and can be mapped over or filtered using user-provided functions.
The \texttt{forall} and \texttt{exists} methods check if a given predicate holds for all (respectively for at least one) element of the set.

\paragraph{Maps.}
Associate keys to values.
Programmers can add key-value pairs, remove keys, and fetch the value that is associated to a certain key.
The \texttt{keys} (resp. \texttt{values}) method returns a set containing all keys (resp. values) contained by the map.
The \texttt{bijective} method checks if there is a one-to-one correspondence between the keys and the values.
Maps support many well-known functional operations;
\texttt{zip} returns a map of tuples containing only the keys that are present in both maps and stores their values in a tuple;
\texttt{combine} returns a map containing \emph{all} entries from both maps, using a user-provided function \texttt{f} to combine values that are present in both maps.

\paragraph{Vectors.}
Represent a sequence of elements which are indexed from \texttt{0} to \texttt{size-1}.
Elements can be written to a certain index which will overwrite the existing value at that index.
One can append a value to the vector which will write that value at index \texttt{size}, thereby, making the vector grow.
Like sets and maps, programmers can map functions over vectors, zip vectors, and check predicates for all or for one element of a vector.

\paragraph{Lists.}
Represent a sequence of elements in a linked list.
Unlike vectors, \texttt{insert} does not overwrite the existing value at that index.
Instead, the existing value at that index and all subsequent values are moved one position to the right.
Elements can also be deleted from a list, making the list shrink.

\section{Automated Verification}\label{sec:auto-verif}

\VeriFx leverages SMT solvers to enable automated verification.
Such solvers try to (automatically) determine whether or not a given formula is satisfiable.
Modern SMT solvers support various specialized theories (for bitvectors, arrays, etc.) and are very powerful if care is taken to encode programs efficiently using these theories.
However, SMT-LIB, a standardized language for SMT solvers\footnote{\url{http://smtlib.cs.uiowa.edu/}}, is low-level and is not meant to be used directly by programmers to verify high-level programs.
Instead, semi-automatic program verification usually involves implementing the program in an Intermediate Verification Language (IVL) which internally compiles to SMT-LIB to discharge the proof obligations using an appropriate SMT solver.
IVLs like Dafny~\cite{dafny}, Spec\#~\cite{specSharp}, and Why3~\cite{why3})
are designed to be general-purpose but this breaks automated verification since
programmers need to specify preconditions and postconditions on methods, loop invariants, etc.

\VeriFx can be seen as a specialized high-level IVL that was %
carefully designed such that every feature has an efficient SMT encoding; %
leaving out features that break automated verification. %
For example, \VeriFx does not support traditional loop statements but instead provides higher order operations (map, filter, etc.) on top of its functional collections.
The resulting language is surprisingly expressive given its automated verification capabilities.

In the remainder of this section we show how \VeriFx compiles programs to SMT and derives proof obligations that can be discharged automatically by SMT solvers.
Afterwards, we explain how \VeriFx leverages a specialised theory of arrays to efficiently encode its functional collections.
Due to space constraints, \cref{app:compilation-example} exemplifies these compilation rules using a concrete example.

\subsection{Core SMT}\label{sec:core-smt-syntax}

The semantics of \VeriFx are defined using translation functions from VeriFx to Core SMT, %
a reduced version of SMT that suffices to verify \VeriFx programs.
\Cref{fig:smt-constructs} defines the syntax of Core SMT.
The metavariable
$\sortName$ ranges over user-declared sorts\footnote{The literature on SMT solvers uses the term ``sort'' to refer to types and type constructors.};
$\adtName$ ranges over names of algebraic data types (ADTs); %
$\ctorName$ ranges over ADT constructor names;
$\typeParam$ ranges over type variables;
$\field$ ranges over field names;
$\funName$ ranges over function names;
$\type$ ranges over types;
$\param$ ranges over variable names; %
$\expr$ ranges over expressions;
and $\intName$ ranges over integers.
Valid types include integers, strings, booleans, arrays, ADTs $\customType{\adtName}{\zeroOrMore{\type}}$, and user-declared sorts $\customType{\sortName}{\zeroOrMore{\type}}$.
Arrays are \emph{total} and map values of 
the key types to a value of the element type.
Arrays can be multidimensional and map several keys to a value. %

\begin{figure}[tbp]
\centering
  \centering
  $$
  \small{\begin{array}{@{}r@{\,}c@{\,}l @{}r@{\,}c@{\,}l}
  \type & ::= & \intsmt \, \mid \, \stringsmt \, \mid \, \boolsmt  &
  \qquad
    \mathit{G} & ::= & \adtsmt{\adtName}{\typeParams}{\oneOrMore{ \ctorsmt{\ctorName}{\argsmt{\zeroOrMore{\field}} {\zeroOrMore{\type}}} }} \\ %
            & \mid &     \arrayType{\oneOrMore{\type}}{\type} \, \mid \,
				\customType{\adtName}{\zeroOrMore{\type}} \, \mid \,
				\customType{\sortName}{\zeroOrMore{\type}} &
	\qquad     \expr & ::= & \readsmtt{\exprsmt}{\oneOrMore{\exprsmt}} \, \mid \, %
			    \writesmtt{\exprsmt}{\oneOrMore{\exprsmt}}{\exprsmt} \, \mid \, %
				\lambdasmt{ \argsmt{\oneOrMore{\param}}{\oneOrMore{\type}} } { \exprsmt } \\  
  \mathit{C} & ::= & \constsmt{\varName}{\type} &
  \qquad 
  	& \mid & \forallsmt{ \argsmt{\oneOrMore{\param}} {\oneOrMore{\type}} }{\exprsmt} \, \mid \,
	 \existssmt{ \argsmt{\oneOrMore{\param}} {\oneOrMore{\type}} }{\exprsmt} \, \mid \, %
  				\ldots
\\  	
  \mathit{D} & ::= & \sortsmt{\sortName}{\intName} &
  \qquad
    \assertion & ::= & \assertsmt{\exprsmt} \\ %
  \FunDef & ::= & \funDefsmt{\funName}{\typeParams}{\argsmt{\zeroOrMore{\param}} { \zeroOrMore{\type}}}{\type}{\exprsmt} &
  \qquad
   H & ::= & \checksmt()
  \end{array}}
  $$
  \vspace{-3mm}
  \captionof{figure}{Core SMT syntax.}
  \label{fig:smt-constructs}
\end{figure}

Core SMT programs consist of one or more statements which can be the declaration of a constant or sort, assertions, the definition of a function or ADT, or a call to $\checksmt$. %
Constant declarations take a name and a type.
Sort declarations take a name and a non-negative number $i$ representing their arity, \ie how many type parameters the sort takes.
Declared constants and sorts are \emph{uninterpreted} and the SMT solver is free to assign any valid interpretation. %
Assertions are boolean formulas that constrain the possible interpretations of the program, \eg $\assertsmt{\texttt{age} >= 18}$.

Function definitions consist of a name $\funName$, optional type parameters $\typeParams$, formal parameters $\argsmt{\zeroOrMore{\param}} { \zeroOrMore{\type}}$, a return type $\type$, and a body containing an expression $\expr$. %
Valid expressions include array accesses $\readsmt{\exprsmt}{\oneOrMore{\exprsmt}}$, array updates $\writesmt{\exprsmt}{\oneOrMore{\exprsmt}}{\exprsmt}$, anonymous functions, quantified formulas, etc\footnote{The complete list of expressions is described in~\cref{app:smt-expressions} as part of the additional material.}.
Updating an array returns a modified copy of the array.
It is important to note that arrays are total and that anonymous functions define an array from the argument types to the return type.
For example, $\lambdasmt{\argsmt{x}{\intsmt}, \, \argsmt{y}{\intsmt}}{x+y}$ defines an $\arrayType{\intsmt, \intsmt}{\intsmt}$ that maps two integers to their sum.
Since arrays are first-class values in SMT, it follows that lambdas are also first-class.

ADT definitions consist of a name $\adtName$, optional type parameters $\typeParams$, and one or more constructors.
Every constructor has a name $\ctorName$ and optionally defines fields with a name $\field$ and a type $\type$.
Constructors are invoked like regular functions and return an instance of the data type.

The decision procedure ($\checksmt$) checks the satisfiability of the SMT program. %
If the program's assertions are satisfiable, $\checksmt$ returns a concrete model,
\ie an interpretation of the constants and sorts that satisfies the assertions.
A property $p$ can be proven by showing that the negation $\neg p$ is unsatisfiable, \ie that no counterexample exists.

Note that our Core SMT language includes lambdas and polymorphic functions which are not part of SMT-LIB v2.6.
Nevertheless, they are described in the preliminary proposal for SMT-LIB v3.0\footnote{\url{http://smtlib.cs.uiowa.edu/version3.shtml}} and Z3 already supports lambdas.
For the time being, \VeriFx monomorphizes polymorphic functions when they are compiled to Core SMT. %
For example, given a polymorphic identity function \texttt{id<X> :: X -> X}, \VeriFx creates a monomorphic version \texttt{id\_int :: int -> int} when encountering a call to \texttt{id} with an integer argument.

\subsection{Compiling \VeriFx to SMT}\label{sec:transformations}

Similarly to Dafny~\cite{dafny}, we describe the semantics of \VeriFx by means of translation functions that compile \VeriFx programs to Core SMT.
Types are translated by the $\convtype{}$ function: %
$$
\small{\begin{array}{lcllcllcl}
	\convtype{\boolv} & = & \boolsmt  & \quad
		\convtype{\intv} & = & \intsmt  & \quad
		\convtype{\stringv} & =  &\stringsmt \\[2em]
    	\convtype{\vfx{\customType{\className}{\zeroOrMore{T}}}}  & = & \customType{\className} {\zeroOrMore{\convtype{\vfx{T}}}}  & \quad
    
	\convtype{\vfx{\customType{\enumName}{\zeroOrMore{T}}}}  & = & \customType{\enumName}{\zeroOrMore{\convtype{\vfx{T}}}} & \quad

	\convtype{\funTypeV{\oneOrMore{\vfx{\type}}}{\vfx{\typeP}}}  & =  & \arrayType{\oneOrMore{\convtype{\vfx{\type}}}}{\convtype{\vfx{\typeP}}}
\end{array}}
$$

Primitive types are translated to the corresponding primitive type in Core SMT.
Class types and enumeration types keep the same type name and their type arguments are translated recursively $\zeroOrMore{\convtype{\vfx{T}}}$.
Functions are encoded as arrays from the argument types to the return type.
Trait types do not exist in the compiled SMT program because traits are compiled away by \VeriFx,
\ie only the types of the classes that implement the trait exist in the SMT program.

We now take a look at the translation function $\convdef{}$ which compiles \VeriFx's main constructs: enumerations, classes, and objects.
Enumerations are encoded as ADTs:
\small{\[ \convdef{\adt{\enumName}{\typeParams}{\oneOrMore{\constructor{\ctorName}{\zeroOrMore{\field} : \zeroOrMore{\type}}}}} \ =\ \adtsmt {\enumName} { \typeParams } { \oneOrMore{\ctorsmt {\ctorName} { \argsmt{\zeroOrMore{\field}}{\zeroOrMore{\convtype{\type}}} }} } \]}
\normalsize
For every enumeration an ADT is constructed with the same name, type parameters, and constructors.
The types of the fields are translated recursively.

Classes are encoded as ADTs with one constructor and class methods become functions:
$$
\small{\begin{array}{l}
\convdef{\classv{\className}{\typeParams}{\zeroOrMore{\field} : \zeroOrMore{ \type }}{\methods} \, \mathtt{extends} \, \customType{\traitName}{\zeroOrMore{\typeP}}}  \ = \\[0.1em]
	\quad \adtsmt {\className} {\typeParams} { \, \ctorsmt { \ctorName } { \argsmt{\zeroOrMore{\field}}{\zeroOrMore{\convtype{\type}}} } \, }  \, \seq %
	\, \zeroOrMore{\convmethod{\className, \typeParams, \method}}
	\seq \, \zeroOrMore{\convmethod{\className, \typeParams, \method'[\zeroOrMore{\typeP}/\typeParamsY] }} \\
	\quad \mbox{where } \ctorName = \concat{\className}{"\_ctor"} \mbox{ and } \traitName \mbox{ is defined as } \trait{\traitName}{\typeParamsY}{\zeroOrMore{\method'} \seq \ldots}
\\\\[-8pt]
\convmethod{\className , \typeParams, \methodv{\methodName}{\typeParamsY}{\zeroOrMore{\varName} : \zeroOrMore{ \type }}{T_r}{\expr}} \ = %
	\funDefsmt {\funName}
	{ \typeParams, \typeParamsY }
	{ \argsmt{this}{\customType{\className}{\typeParams}}, \argsmt{\zeroOrMore{\varName}}{\zeroOrMore{\convtype{\type}}} } {\convtype{T_r}} {\convexp{\expr}}\\
	\quad \mbox{where } \funName = \concatt{\className}{"\_"}{\methodName}
\end{array}}
$$

The ADT keeps the name of the class and its type parameters, and defines one constructor containing the class' fields.
Since the name of the constructor must differ from the ADT's name, the compiler defines a unique name $\ctorName$ which is the name of the class followed by ``\_ctor''.
The class methods $\zeroOrMore{\method}$ are compiled to regular functions by the $\convmethod{}$ function.
Furthermore, the class inherits all concrete methods $\zeroOrMore{\method'}$ that are defined by its super trait and are not overridden by itself.
This requires substituting the trait's type parameters $\typeParamsY$ by the concrete type arguments $\zeroOrMore{\typeP}$ provided by the class. %
As such, traits are compiled away and do not exist in the transpiled SMT program.

For every method, a function is created with a unique name $\funName$ that is the name of the class followed by an underscore and the name of the method.
In the argument list, the body, and the return type of a method, programmers can refer to type parameters of the class and type parameters of the method.
Therefore, the compiled SMT function takes both the class' type parameters $\typeParams$ and the method's type parameters $\typeParamsY$.
Without loss of generality we assume that a method's type parameters do not override the class' type parameters which can be achieved through $\alpha$-conversion.
The method's parameters become parameters of the function.
In addition, the function takes an additional parameter $this$ referring to the receiver of the method call which should be of the class' type.
The types of the parameters and the return type are translated using function $\convtype{}$.
The body of the method must be a well-typed expression.
Expressions are translated by the $\convexp{}$ function: %
$$
\small{\begin{array}{lcl lcl}
\convexp{\varName} & = &\varName &
	\qquad
	\convexp{\new{\customType{\ctorName}{\zeroOrMore{\type}}}{\exprs}} & = & 
	\callWithTypessmt{ \ctorName } { \zeroOrMore{\convtype{\type}} } { \zeroOrMore{\convexp{\expr}} }
\\
\convexp{\varvv{\varName}{T}{e_1}{e_2}}  & = & \letsmt{\varName}{\convexp{e_1}}{ \convexp{e_2}} &
\qquad
\convexp{\accessv{\expr}{\field}} & = & \accesssmt{\convexp{\expr}}{\field} \\
\convexp{\lambdav{\oneOrMore{\varName} : \oneOrMore{\type}} {\expr}}  & =  &
	\lambdasmt{ \argsmt{\oneOrMore{\varName}}{\oneOrMore{\convtype{\type}}} } { \convexp{\expr} } &
	\qquad
\convexp{\invoke{\expr_1}{\methodName}{\zeroOrMore{\type}}{\exprs}} & = &
	  \callWithTypessmt{ \methodName' } { \zeroOrMore{\convtype{\typeP}} , \, \zeroOrMore{\convtype{\type}} } { \convexp{\expr_1}, \, \zeroOrMore{\convexp{\expr}} }
	 \\
\convexp{\callv{\expr_1}{\zeroOrMore{\expr_2}}}  & = & 
	\readsmt{ \convexp{\expr_1} } { \zeroOrMore{\convexp{\expr_2}} } & 
	 \multicolumn{3}{l}{\qquad \quad \mbox{where } \mathit{typeof}(e_1) = \className \langle \zeroOrMore{\typeP} \rangle}\\
\convexp{\new{\customType{\className}{\zeroOrMore{\type}}}{\exprs}} & = &
	\callWithTypessmt{ \className' } { \zeroOrMore{\convtype{\type}} } { \zeroOrMore{\convexp{\expr}} } &

	\multicolumn{3}{l}{\qquad \quad \mbox{ and } \methodName' = 
	\concatt{\className}{"\_"}{\methodName} \mbox{ and } \zeroOrMore{\typeP} \cap \zeroOrMore{\type} = \emptyset}
	\\
	\multicolumn{3}{l}{\quad \mbox{where } \className' = \concat{\className}{"\_ctor"} } 
\end{array}}
$$

Primitive values, variable references, and parameter references remain unchanged in Core SMT.
The definition of an immutable variable is translated to a let expression.
Anonymous functions remain anonymous functions in Core SMT, the type of the parameters and the body are compiled recursively.
Remember that anonymous functions in SMT define (multidimensional) arrays from one or more arguments to the function's return value. 
Hence, function calls are translated to array accesses.
To instantiate a class or ADT, the compiler calls the data type's constructor function.
For classes, the constructor's name is the name of the class followed by ``\_ctor''.
To access a field, the compiler translates the expression and accesses the field on the translated expression.
To invoke a method $\methodName$ on an object $e_1$ the compiler calls the corresponding function $\methodName'$ which by convention is the name of the class followed by an underscore and the name of the method.
Recall that the function takes both the class' type arguments $\types$ and the method's type arguments $\zeroOrMore{\typeP}$ as well as an additional argument $e_1$ which is the receiver of the call.
The complete set of compilation rules for expressions is provided in~\cref{app:compile-expressions} as part of the additional material.

Finally, objects are singletons that can define methods and proofs, and are compiled as follows:
$$
\small{\begin{array}{l}
\convdef{ \mathtt{object} \, \objectName \, \mathtt{extends} \, \customType{\traitName}{\types} \, \{ \, \methods \seq \zeroOrMore{\prooff} \, \} } \ = \ \\ \quad
  \convdef{ \mathtt{class} \, \objectName' ( \, ) \, \{ \methods \} \, \mathtt{extends} \, \customType{\traitName}{\types} } \seq \, %
  \constsmt{\objectName}{\objectName'} \seq \,
  \assertsmt{\objectName == \objectName'()} \seq \,

  \convdef{\zeroOrMore{\prooff}} %
\end{array}}
$$
The object is compiled to a regular class with a fresh name $\objectName'$.
Then, a single instance of that class is created and assigned to a constant named after the object $O$.
The proofs defined by the object are compiled to functions.
How to translate proofs into functions is the subject of the next section. %

\subsection{Deriving Proof Obligations}\label{sec:compiling-proofs}

We previously verified a 2PSet CRDT using \VeriFx's CRDT library which
internally uses our novel proof construct to define the necessary correctness properties (discussed later in~\cref{sec:rdt-libs}).
However, programmers can also define custom proofs, for instance to verify data invariants. %

We now explain how proof obligations are derived from user-defined proofs in \VeriFx programs.
Proofs are compiled to regular functions without arguments.
The name and type parameters remain unchanged and the body of the proof is compiled and becomes the function's body.
Proofs always return a boolean since the body is a logical formula whose satisfiability must be checked.
\[
\small{\begin{array}{l}
\convdef{\proofv{\proofName}{\typeParams}{\expr}} \ =\
\funDefsmt {\proofName} {\typeParams} {} {\boolsmt} {\convexp{\expr}} 
\end{array}} 
\]

To check if the property described by a proof holds, the negation of the proof must be unsatisfiable.
In other words, if no counterexample exists it constitutes a proof that the property is correct.
A (polymorphic) proof called $\proofName$ with zero or more type parameters $i$ is checked as follows: 
$$
\small{\begin{array}{lcl}
\checkproof{\proofName}{i} & = & 
    \sortsmt{\sortName_1}{0} \; ; \; \ldots \; ; \; \sortsmt{\sortName_i}{0} \; \seq \; %
    \assertsmt{ \neg \callWithTypessmt{\proofName}{\sortName_1, \ldots, \sortName_i}{} } \; \seq \;
    \checksmt() == \smt{UNSAT} %
\end{array}}
$$

For every type parameter an \emph{uninterpreted} sort is declared. %
Then, the proof function is called with those sorts as type arguments and we check that the negation is unsatisfiable.
If the negation is unsatisfiable, the (polymorphic) proof holds for all possible instantiations of its type parameters. %
The underlying SMT solver can generate an actual proof which 
could be reconstructed by proof assistants as shown by \citet{reconstructBitVectors, bohme2010fast}.

\subsection{Encoding Functional Collections Efficiently in SMT}

Some IVLs feature collections with rich APIs (\eg Why3~\cite{why3}) but encode operations on these collections recursively. %
Traditional SMT solvers
fail to verify recursive definitions automatically because they require inductive proofs,
which is beyond the capabilities of most solvers.
However, many SMT solvers support specialised array theories. %
A key insight of this paper consists of
efficiently encoding the collections and their operations using the Combinatory Array Logic (CAL)~\cite{z3ArrayTheory} which is decidable. %
As a result, \VeriFx can automatically verify RDTs that
are built by arbitrary compositions of functional collections.
In the remainder of this section we describe the encoding of the different functional collections using this array logic.

\subsubsection{Set Encoding}

Sets are encoded as arrays from the element type to a boolean type that indicates whether the element is in the set:
\[ 
\small{\begin{array}{l}
\convtype{\setv{\type}} = \arrayType{\convtype{\vfx{\type}}}{\boolsmt} 
\end{array}}
\]
An empty set corresponds to an array containing false for every element.
We can create such an array by defining a lambda that ignores its argument and always returns false:
\[ 
\small{\begin{array}{l}
\convexp{\new{\vfx{\setv{\type}}}{}} =  \lambdasmt{ \argsmt{\param}{\convtype{\type}} }{\smt{false}} 
\end{array}}
\]
Operations on sets are compiled as follows: %
$$
\begin{small}
\begin{array}{lcl}
 \convset{e_1.\mathit{add}(e_2)}  = \writesmt {\convexp{e_1}} { \convexp{e_2} } { \smt{true} }  %
& \convset{e_1.\mathit{remove}(e_2)} = \writesmt {\convexp{e_1}} { \convexp{e_2} } { \smt{false} } %
& \convset{e_1.\mathit{contains}(e_2)} = \readsmt {\convexp{e_1}} {\convexp{e_2}} \\
\\[-10pt]
\multicolumn{3}{l}{\convset{e_1.\mathit{filter(e_2)}} =
   \lambdasmt{ \argsmt{\varName}{\convtype{T}} } { \readsmt{\convexp{e_1}}{\varName} \wedge \readsmt{\convexp{e_2}}{\varName} }%
   \quad \mbox{where } \mathit{typeof}(e_1) =  \mathtt{Set} \langle T \rangle \wedge \mathit{typeof}(\expr_2) = 
	\funTypeV{\type}{\boolv}} \\
\\[-10pt]

\multicolumn{3}{l}{\convset{\expr_1.\mathit{map(\expr_2)}} =
    \lambdasmt{ \argsmt{\paramY}{\convtype{\typeP}} } {  
        \existssmt{\argsmt{\varName}{\convtype{\type}}} { %
           \readsmt{\convexp{\expr_1}}{\varName} \wedge \readsmt{\convexp{\expr_2}}{\varName} = \paramY} }}\\
 \multicolumn{3}{l}{\qquad  \mbox{where } \mathit{typeof}(\expr_1) =  \mathtt{Set} \langle \type \rangle \wedge \mathit{typeof}(\expr_2) =  \type \rightarrow \typeP}
\end{array}
\end{small}
$$
An element $e_2$ is added to a set $e_1$ by setting the entry for $e_2$ in the array that results from transforming $e_1$ to true.
Similarly, an element is removed by changing its entry in the array to false.
An element is  in the set if its entry is true.
A set $e_1$ containing elements of type $\type$ can be filtered such that only the elements that fulfil a given predicate $\expr_2 : \funTypeV{\type}{\boolv}$ are retained.
Calls to $\mathit{filter}$ are compiled to a lambda that defines a set (\ie an array from elements to booleans) containing only the elements $\param$ that are in the original set $\expr_1$ (\ie $\readsmt{\convexp{\expr_1}}{\param}$) and fulfil predicate $\expr_2$ (\ie $\readsmt{\convexp{\expr_2}}{\param}$).
Similarly, a function $e_2 : \funTypeV{\type}{\typeP}$ can be mapped over a set $e_1$ of $\type$s, yielding a set of $\typeP$s. Calls to $map$ are compiled to a lambda that defines a set containing elements $\paramY$ of type $\convtype{\typeP}$ such that an element $\param$ exists that is in the original set $\expr_1$ (\ie $\readsmt{\convexp{\expr_1}}{\param}$) and maps to $\paramY$ (\ie $\readsmt{\convexp{\expr_2}}{\param} = \paramY$).
The remaining methods are similar and are described in \cref{app:full-set-implementation} as part of the additional material. %

\subsubsection{Map Encoding}\label{sec:map-encoding}

Maps are encoded as arrays from the key type to an optional value:
\[ 
\small{\begin{array}{l}
\convtype{\mapv{\type}{\typeP}} = \arrayType{ \convtype{\vfx{\type}} } { \optionType{\convtype{\vfx{\typeP}}} } 
\end{array}}
\]
Optional values indicate the presence or absence of a value for a certain key. %
The option type is defined as an ADT with two constructors:
$\smt{Some}(value)$ which holds a value and $\smt{None}()$ indicating the absence of a value.
An empty map corresponds to an array containing $\smt{None}()$ for every key and is created by a lambda that returns $\smt{None}()$ for every key:
\[ 
\small{\begin{array}{l}
\convexp{\new{\vfx{\mapv{\type}{\typeP}}}{}} = \lambdasmt{\argsmt{\param} {\convtype{\type}}}{\none{\convtype{\typeP}}} 
\end{array}}
\]

Operations on maps are compiled as follows:
$$
\begin{small}
\begin{array}{ll}
\!\!\!\convmap{e_m.\mathit{add}(e_k, e_v)} = %
\writesmt{\convexp{e_m}} { \convexp{e_k} } { \some { \convexp{e_v} }} 

&\!\!\convmap{e_m.\mathit{remove}(e_k)} = %
\writesmt{\convexp{e_m}} { \convexp{e_k} } { \none{\convtype{V}} }\\
\\[-10pt]
\!\!\!\convmap{e_m.\mathit{contains}(e_k)} = %
\readsmt {\convexp{e_m}} { \convexp{e_k} } \neq \none{\convtype{V}}\
&\!\!\convmap{e_m.\mathit{get}(e_k)} = %
\accesssmt {\readsmt {\convexp{e_m}} { \convexp{e_k} }} { \mathit{value} }\\
\\[-10pt]
\multicolumn{2}{l}{\!\!\!\convmap{e_m.\mathit{getOrElse}(e_k, e_v)} = %
 \ifsmt{ \readsmt {\convexp{e_m}} { \convexp{e_k} } = \none{\convtype{V}} }
{  \convexp{e_v} }
{  \accesssmt {\readsmt {\convexp{e_m}} { \convexp{e_k} }} {\mathit{value}} }}
\end{array}
\end{small}
$$
A key-value pair $e_k \mapsto e_v$ is added to a map $e_m$ by updating the entry for the compiled key $\convexp{e_k}$ in the compiled array $\convexp{e_m}$ with the compiled value, $\some { \convexp{e_v} }$.
A key $e_k$ is removed from a map $e_m$ by updating the corresponding entry to $\none{\convtype{V}}$, thereby indicating the absence of a value. %
Note that $\smt{None}$ is polymorphic but the type parameter cannot be inferred from the arguments; it is thus passed explicitly. 
A key $e_k$ is present in a map $e_m$ if the value that is associated to the key is not $\none{\convtype{V}}$.
The $\texttt{get}$ method fetches the value that is associated to a key $e_k$ in a map $e_m$.
To this end, the compiled key $\convexp{e_k}$ is accessed in the compiled map $\convexp{e_m}$ and the value it holds is then fetched by accessing the $\texttt{value}$ field of the $\smt{Some}$ constructor.
Even though the entry that is read from the array is an option type (\ie a $\smt{None}$ or a $\smt{Some}$) we can access the $\texttt{value}$ field because the interpretation of $\texttt{value}$ is underspecified in SMT.
If the entry is a $\smt{None}$, the SMT solver can assign any interpretation to the $\texttt{value}$ field.
Hence, %
the $\texttt{get}$ method on maps 
 should only be called if the key is known to be present in the map, \eg after calling $\texttt{contains}$.
\VeriFx also features a safe variant, called $\texttt{getOrElse}$, 
which returns a default value if the key is not present.

We now show how a selection of advanced map operations are compiled:
\small
\begin{align*}
  &\convmap{e_m.\mathit{keys}()} =
  \lambdasmt{ \argsmt{\var}{\convtype{K}} } { \readsmt{\convexp{e_m}}{\var} \neq \none{\convtype{V}} }
\quad \mbox{where } \mathit{typeof}(e_m) = \mathtt{Map} \langle K, V \rangle \\[-0.3em]
  &\convmap{e_m.\mathit{map}(e_f)} =
 \lambdasmt{ \argsmt{\var}{\convtype{K}} }{\ 
\ifsmt{\readsmt{\convexp{e_m}} {\var} \neq \none{\convtype{V}} }
    { \\[-0.3em] & \qquad \qquad \qquad \qquad \qquad \qquad \qquad \; \; \some{ \readsmt{\convexp{e_f}} {\var, \accesssmt{\readsmt{\convexp{e_m}} {\var}} {\mathit{value}} } } }
    { %
      \, \none{\convtype{W}} } }\\[-0.3em]
    & \quad \mbox{where } \mathit{typeof}(e_m) = \mathtt{Map} \langle K, V \rangle
\mbox{ and } \mathit{typeof}(e_f) = (K, V) \rightarrow W \\[-0.3em]
\end{align*}
\normalsize
The \texttt{keys} method returns a set containing only the keys that are present in the map.
Calls to \texttt{keys} on a map $e_m$ of type $\mapv{K}{V}$ are compiled to a lambda which defines a set of keys $x$ of the compiled key type $\convtype{K}$ such that a key is present in the set iff it is present in the compiled map, \ie $ \readsmt{\convexp{e_m}}{x} \neq \none{\convtype{V}} $.
Mapping a function $e_f$ over the key-value pairs of a map $e_m$ is encoded as a lambda that defines an array containing only the keys that are present in the compiled map $\convexp{e_m}$ and whose values are the result of applying $e_f$ on the key and its associated value, \ie $\some{ \readsmt{\convexp{e_f}} {x, \accesssmt{\readsmt{\convexp{e_m}} {x}} {\mathit{value}} } }$.
The remaining operations (\cf \cref{fig:verifx-collections}) are encoded similarly and are described in \cref{app:full-map-implementation} as part of the additional material.

\subsubsection{Vectors and Lists.}
The encoding of sets and maps is very useful to build new data structures in \VeriFx without having to encode them manually in SMT.
For example, vectors and lists are implemented on top of maps.
Internally, they map indices between $0$ and $size - 1$ to their value, and provide a traditional interface on top (\cf \cref{fig:verifx-collections}).
Note that this encoding of vectors and lists on top of maps is only used when verifying proofs in SMT;
when compiling to a target language (\eg Scala or JavaScript), \VeriFx leverages the language's built-in vector and list data structures.

\section{Libraries for Implementing and Verifying Replicated Data Types}\label{sec:rdt-libs}

To simplify the development of distributed systems that use replicated data, we build two libraries for implementing and automatically verifying RDTs that use the CRDT or OT approach.
We first discuss the implementation of a general execution model for CRDTs and its verification library in \VeriFx.
Afterwards, we present a library for implementing RDTs using OT and verifying the transformation functions.
\VeriFx is not limited to these two families of RDTs;
programmers can build custom libraries for implementing and verifying other abstractions or families of RDTs.
This section describes the core of the libraries. Their implementation will be in the artifact.

\subsection{CRDT Library}\label{sec:crdt-lib}

CRDTs guarantee \emph{strong eventual consistency} (SEC), a consistency model that strengthens eventual consistency with the \emph{strong convergence} property which requires replicas that received the same updates, possibly in a different order, to be in the same state.
\VeriFx's CRDT library supports several families of CRDTs, including state-based~\cite{CRDTs}, op-based~\cite{CRDTs}, and pure op-based CRDTs~\cite{Baquero17}.

\subsubsection{State-based CRDTs.}

State-based CRDTs (CvRDTs for short) periodically broadcast their state to all replicas and merge incoming states by computing the least upper bound (LUB) of the incoming state and their own state.
\Citet{CRDTs} showed that CvRDTs converge if the merge function $\lub$ is idempotent, commutative, and associative.
We define these properties as follows:
\begin{description}
\item [Idempotent:]
  $\forall x \in \Statee: \reachable{x} \implies x \equiv x \lub x$
\item [Commutative:]
  $\forall x, y \in \Statee: \reachable{x} \, \land \, \reachable{y} \land \compat{x}{y} \\ \implies
 (x \lub y \equiv y \lub x) \, \land \, \reachable{x \lub y}$
\item [Associative:]
  $\forall x, y, z \in \Statee: \reachable{x} \, \land \, \reachable{y} \, \land \, \reachable{z} \, \land \\ \compat{x}{y} \, \land \, \compat{x}{z} \, \land \, \compat{y}{z} \\ \implies ((x \lub y) \lub z \equiv x \lub (y \lub z)) \, \land \, \reachable{(x \lub y) \lub z}$
\end{description}
$\Statee$ denotes the set of all states.
A state is $reachable$ if it can be reached starting from the initial state and applying only supported operations.
Two states are $compatible$ if they represent different replicas of the same CRDT object\footnote{This definition of compatibility allows replicas to keep unique information, \eg to generate unique tags.}.
As explained in \cref{sec:motiv-design-and-impl}, state equivalence is defined in terms of $\latticeSmaller$ on the lattice: $S \equiv T \iff S \latticeSmaller T \land T \latticeSmaller S$.

\VeriFx's CRDT library provides traits for the implementation and verification of CvRDTs.
\Cref{lst:CvRDT-exec-model-impl} shows the \texttt{CvRDT} trait %
that was used in \cref{lst:2PSet-VFx} to implement the \texttt{TwoPSet} CRDT.
Every state-based CRDT that extends the \texttt{CvRDT} trait must provide a type argument which is the actual type of the CRDT and provide an implementation for the \texttt{merge} and \texttt{compare} methods.
By default, all states are considered reachable and compatible, and state equivalence is defined in terms of \texttt{compare}.
These methods can be overridden by the concrete CRDT that implements the trait.

\Cref{lst:CvRDT-proof-verif-impl} shows the \texttt{CvRDTProof} trait which is used to verify CvRDT implementations.
This trait defines one type parameter \texttt{T} that must be a CvRDT type and defines proofs to check that its merge function adheres to the aforementioned properties (\ie is idempotent, commutative, and associative).
It also defines an additional proof \texttt{equalityCheck} which checks if the notion of state equivalence that is derived from \texttt{compare} corresponds to structural equality.
As shown in \cref{sec:motivation}, objects can extend this \texttt{CvRDTProof} trait to inherit automated correctness proofs for the given CRDT type.
Note that the trait's type parameter \texttt{T} expects a concrete CvRDT type (\eg \texttt{PNCounter}) and will not work for polymorphic CvRDTs (\eg \texttt{ORSet}) because those are type constructors.
Instead, the CRDT library provides additional \texttt{CvRDTProof1}, \texttt{CvRDTProof2}, and \texttt{CvRDTProof3} traits to verify polymorphic CvRDTs that expect 1, 2, or 3 type arguments respectively. %
For example, the \texttt{TwoPSet[V]} from \cref{sec:motivation} is polymorphic in the type of values it stores; the \texttt{TwoPSetProof} object thus extended the \texttt{CvRDTProof1} trait because the \texttt{TwoPSet} expects one type argument. %

\begin{figure*}
\centering
\begin{minipage}[t]{.37\textwidth}
\begin{lstlisting}[style=VFxStyle, caption={Trait for the implementation of CvRDTs in \VeriFx.}, label=lst:CvRDT-exec-model-impl, basicstyle={\scriptsize\ttfamily}]
trait CvRDT[T <: CvRDT[T]] {
  def merge(that: T): T
  def compare(that: T): Boolean
  def reachable(): Boolean = true
  def compatible(that: T): Boolean = true
  def equals(that: T): Boolean = {
    this.asInstanceOf[T].compare(that) &&
    that.compare(this.asInstanceOf[T])
  }
}
\end{lstlisting}
\end{minipage}
\hfill 
\begin{minipage}[t]{.57\textwidth}
\begin{lstlisting}[style=VFxStyle, caption={Trait for the verification of CvRDTs in \VeriFx. The arrow function \texttt{=>:} implements logical implication.}, label=lst:CvRDT-proof-verif-impl, basicstyle={\scriptsize\ttfamily}]
trait CvRDTProof[T <: CvRDT[T]] {
  proof mergeIdempotent {
    forall (x: T) { x.reachable() =>: x.merge(x).equals(x) } }
  proof mergeCommutative {
    forall (x: T, y: T) {
      (x.reachable() && y.reachable() && x.compatible(y)) =>:
      (x.merge(y).equals(y.merge(x)) && x.merge(y).reachable())}}
  proof mergeAssociative {
    forall (x: T, y: T, z: T) {
      (x.reachable() && y.reachable() && z.reachable() &&
       x.compatible(y) && x.compatible(z) && y.compatible(z)) =>:
        (x.merge(y).merge(z).equals(x.merge(y.merge(z))) &&
         x.merge(y).merge(z).reachable()) } }
  proof equalityCheck {
    forall (x: T, y: T) { x.equals(y) == (x == y) } } }
\end{lstlisting}
\end{minipage}
\end{figure*}

\subsubsection{Op-based CRDTs.}\label{sec:CmRDT-lib}

Op-based CRDTs (CmRDTs for short) execute update operations in two phases, called \emph{prepare} and \emph{effect}. The prepare phase executes locally at the source replica (only if its source precondition holds) and prepares a message to be broadcast\footnote{While some CmRDT designs do not require causal delivery, the overall model assumes reliable causal broadcast.} to all replicas (including itself).
The effect phase applies such incoming messages and updates the state (only if its downstream precondition holds, otherwise the message is ignored).

\Citet{CRDTs} and \citet{IsabelleCRDTs} have shown that CmRDTs guarantee SEC if all concurrent operations commute. %
Hence, for any CmRDT it suffices to show that all pairs of concurrent operations commute.
Formally, for any operation $o_1$ that is enabled by some reachable replica state $s_1$ (\ie $o_1$'s source precondition holds in $s_1$) and any operation $o_2$ that is enabled by some reachable replica state $s_2$, if these operations can be concurrent, and $s_1$, $s_2$, and $s_3$ are compatible replica states, then we must show that on any reachable replica state $s_3$ the operations commute and the intermediate and resulting states are all reachable:
\small
\begin{align*}
&\forall s_1, s_2, s_3 \in \Statee, \forall o_1, o_2 \in \Op:%
\reachable{s_1} \, \land \, \reachable{s_2} \, \land \, \reachable{s_3} \, \land \\[-2pt]
&\quad \srcPrecond{o_1}{s_1} \, \land \, \srcPrecond{o_2}{s_2} \, \land \,
 canConcur(o_1, o_2) \, \land \, \\[-2pt]
&\quad \compat{s_1}{s_2} \, \land \, \compat{s_1}{s_3} \, \land \, \compat{s_2}{s_3} \\[-2pt]
&\quad \implies \applyIfEnabled{o_2}{\applyIfEnabled{o_1}{s_3}} \equiv \applyIfEnabled{o_1}{\applyIfEnabled{o_2}{s_3}} \, \land \, \reachable{\applyIfEnabled{o_1}{s_3}} \, \land  \,\reachable{\applyIfEnabled{o_2}{s_3}} \, \land \,\reachable{\applyIfEnabled{o_1}{\applyIfEnabled{o_2}{s_3}}}
\end{align*}
\normalsize
We use the notation $\applyIfEnabled{o}{s}$ to denote the application of an operation $o$ on state $s$ if its downstream precondition holds, otherwise it returns the state unchanged.

\Cref{lst:CmRDT-exec-model-impl} shows the \texttt{CmRDT} trait that must be extended by op-based CRDTs with concrete type arguments for the supported operations, exchanged messages, and the CRDT type itself.
Every CRDT that extends the \texttt{CmRDT} trait must implement the \texttt{prepare} and \texttt{effect} methods.
The \texttt{tryEffect} method has a default implementation that applies the operation if its downstream precondition holds, otherwise, it returns the state unchanged.
By default, we assume that all states are reachable, that all operations are enabled at the source and downstream, that all operations can occur concurrently, and that all states are compatible.
For most CmRDTs these settings do not need to be altered but
some CmRDTs make other assumptions which can be encoded by overriding the appropriate method.
For example, in an Observed-Removed Set~\cite{shapiro2011comprehensive}) it is not possible to delete tags that are added concurrently; this can be encoded by overriding \texttt{canConcur}.

Similarly to state-based CRDTs, our CRDT library provides a \texttt{CmRDTProof} trait and several numbered versions %
to verify op-based CRDTs.
These traits define a general proof of correctness that checks that all operations commute based on the previously described formula.

\begin{figure}
\begin{lstlisting}[style=VFxStyle, caption={Polymorphic \texttt{CmRDT} trait to implement op-based CRDTs in \VeriFx.}, label=lst:CmRDT-exec-model-impl,  basicstyle={\scriptsize\ttfamily}]
trait CmRDT[Op, Msg, T <: CmRDT[Op, Msg, T]] {
  def prepare(op: Op): Msg
  def effect(msg: Msg): T
  def tryEffect(msg: Msg): T = if (this.enabledDown(msg)) this.effect(msg)  else this.asInstanceOf[T]
  def reachable(): Boolean = true // by default all states are considered reachable
  def canConcur(x: Msg, y: Msg): Boolean = true // all ops can occur concurrently
  def compatible(that: T): Boolean = true // all states are compatible
  def enabledSrc(op: Op): Boolean = true // no source preconditions by default
  def enabledDown(msg: Msg): Boolean = true // no downstream preconditions by default
  def equals(that: T): Boolean = this == that
}
\end{lstlisting}
\vspace{-4mm}
\end{figure}

\subsubsection{Pure op-based CRDTs.}
Pure op-based CRDTs are a family of op-based CRDTs that exchange only the operations instead of data-type specific messages.
The effect phase stores incoming operations in a partially ordered log of (concurrent) operations.
Queries are computed against the log and operations do not need to commute.
Data-type specific redundancy relations dictate which operations to store in the log and when to remove operations from the log.
\VeriFx's CRDT library provides a \texttt{PureOpBasedCRDT} trait for implementing pure op-based CRDTs.
The implementing CRDT inherits the prepare and effect phase (which is the same for all pure op-based CRDTs) and only needs to provide an implementation of the redundancy relations.
In addition, the library provides a \texttt{PureCRDTProof} trait (and numbered versions for polymorphic CRDTs) which checks that for any state $s$ and any two concurrent operations $x$ and $y$, their effect is the same independent of the order in which they are received. This correctness condition is a simplification of the one for op-based CRDTs as pure op-based CRDTs do not define source or downstream preconditions. %

\subsection{OT Library}\label{sec:ot-library}

The Operational Transformation (OT)~\cite{OT} approach applies operations locally and propagates them asynchronously to the other replicas. Incoming operations are transformed against previously executed concurrent operations such that the modified operation preserves the intended effect.
Operations are functions from state to state: $Op : \Op$ and are transformed using a transformation function $T: Op \times Op \rightarrow Op$.
Thus, $T(o_1, o_2)$ denotes the operation that results from transforming $o_1$ against a previously executed concurrent operation $o_2$.
\Citet{SuleimanOTProof} and~\citet{SunOT} proved that replicas eventually converge if the transformation function satisfies two properties: $\mathit{TP_1}$ and $\mathit{TP_2}$.
Property $\mathit{TP_1}$ states that any two enabled concurrent operations $o_i$ and $o_j$ must commute after transforming them:
\small
      	\[\forall o_i, o_j \in Op, \forall s \in \Statee:
 enabled(o_i, s) \land enabled(o_j, s) \land canConcur(o_i, o_j) \implies
      	  T(o_j, o_i)(o_i(s)) = T(o_i, o_j)(o_j(s))\]
\normalsize
Property $\mathit{TP_2}$ states that given three enabled concurrent operations $o_i$, $o_j$, and $o_k$, the transformation of $o_k$ does not depend on the order in which operations $o_i$ and $o_j$ are transformed:
\small
      \begin{equation*}
      \begin{split}
      &\forall o_i, o_j, o_k \in Op, \forall s \in \Statee:%
      enabled(o_i, s) \, \land \, enabled(o_j, s) \, \land \, enabled(o_k, s) \, \land \, canConcur(o_i, o_j) \, \land \\
      & \quad canConcur(o_j, o_k) \, \land \, canConcur(o_i, o_k) %
      \implies T(T(o_k, o_i), T(o_j, o_i)) = T(T(o_k, o_j), T(o_i, o_j))
      \end{split}
      \end{equation*}
\normalsize
Note that properties $\mathit{TP_1}$ and $\mathit{TP_2}$ only need to hold for states in which the operations can be generated, represented by the relation $enabled : Op \times \Statee \rightarrow \Boolean$, and only if the two operations can occur concurrently, represented by the relation $canConcur : Op \times Op \rightarrow \Boolean$.

\VeriFx provides a library for implementing and verifying RDTs that use operational transformations. %
Programmers can build custom RDTs by extending the \texttt{OT} trait shown in \cref{lst:OT-exec-model-impl}.
Every RDT that extends the \texttt{OT} trait must provide concrete type arguments for the state and operations, and implement the \texttt{transform} and \texttt{apply} methods.
The \texttt{transform} method transforms an incoming operation against a previously executed concurrent operation.
The \texttt{apply} method applies an operation on the state.
By extending this trait, the RDT inherits proofs for $TP_1$ and $TP_2$.
By default, these proofs assume that operations are always enabled and that all operations can occur concurrently.
If this is not the case, the RDT can override the \texttt{enabled} and \texttt{canConcur} methods respectively.

\begin{figure}
\begin{lstlisting}[style=VFxStyle, caption={Polymorphic \texttt{OT} trait to implement and verify RDTs using operational transformation in \VeriFx.}, label=lst:OT-exec-model-impl,  basicstyle={\scriptsize\ttfamily}]
trait OT[State, Op] {
  def transform(x: Op, y: Op): Op
  def apply(state: State, op: Op): State
  def enabled(op: Op, state: State): Boolean = true
  def canConcur(x: Op, y: Op): Boolean = true
  proof TP1 {
    forall (opI: Op, opJ: Op, st: State) {
      (this.enabled(opI, st) && this.enabled(opJ, st) && this.canConcur(opI, opJ)) =>: {
        this.apply(this.apply(st, opI), this.transform(opJ, opI)) ==
        this.apply(this.apply(st, opJ), this.transform(opI, opJ)) } } }
  proof TP2 {
    forall (opI: Op, opJ: Op, opK: Op, st: State) {
      (this.enabled(opI, st) && this.enabled(opJ, st) && this.enabled(opK, st) &&
       this.canConcur(opI, opJ) && this.canConcur(opJ, opK) && this.canConcur(opI, opK)) =>: {
        this.transform(this.transform(opK, opI), this.transform(opJ, opI)) == 
        this.transform(this.transform(opK, opJ), this.transform(opI, opJ)) } } } }
\end{lstlisting}
\vspace{-4mm}
\end{figure}

Although \VeriFx supports the general execution model of OT, most transformation functions described by the literature 
were specifically designed for collaborative text editing. %
They model text documents as a sequence of characters and operations insert or delete characters at a given position in the document.
Every paper thus describes four transformations functions, one for every pair of operations:
insert-insert, insert-delete, delete-insert, delete-delete. %

Likewise, \VeriFx's OT library provides a \texttt{ListOT} trait that models the state as a list of values and supports insertions and deletions.
RDTs extending the \texttt{ListOT} trait need to implement four methods (\texttt{Tii}, \texttt{Tid}, \texttt{Tdi}, \texttt{Tdd}) corresponding to the transformation functions for transforming insertions against insertions (\texttt{Tii}), insertions against deletions (\texttt{Tid}), deletions against insertions (\texttt{Tdi}), and deletions against deletions (\texttt{Tdd}).
The trait provides a default implementation of \texttt{transform} that dispatches to the corresponding transformation function based on the type of operations, and a default implementation of \texttt{apply} that inserts or deletes a value from the underlying list.

\section{Evaluation}

We now evaluate the applicability of \VeriFx to implement and verify RDTs.
Our evaluation is twofold.
First, we implement and verify numerous CRDTs taken from literature as well as some new variants.
Afterwards, we verify well-known operational transformation functions and some unpublished designs.
We will submit an artifact including all implementations and proofs.

All experiments reported in this section were conducted on AWS using an m5.xlarge VM with 2 virtual CPUs and 8 GiB of RAM.
All benchmarks are implemented using JMH~\cite{JMH}, a benchmarking library for the JVM.
We configured JMH to execute 20 warmup iterations followed by 20 measurement iterations for every benchmark.
To avoid run-to-run variance JMH repeats every benchmark in 3 fresh JVM forks, yielding a total of 60 samples per benchmark.

\subsection{Verifying Conflict-free Replicated Data Types}\label{sec:eval-crdts}

We implemented and verified an extensive portfolio comprising 35 CRDTs, taken from literature~\cite{shapiro2011comprehensive, Baquero17, shapiroStateBasedORSet, bieniusa2012optimized, kleppmannMap}.
To the best of our knowledge, we are the first to mechanically verify all CRDTs from~\citet{shapiro2011comprehensive}, the pure op-based CRDTs from~\citet{Baquero17}, and the map CRDT from~\citet{kleppmannMap}.

\newcommand{\opBased}{O}
\newcommand{\stateBased}{S}
\newcommand{\pure}{P}
\renewcommand\cellalign{cl}

\newcommand{\counterRow}{}%
\newcommand{\flagRow}{\rowcolor{gray!10}}
\newcommand{\registerRow}{} %
\newcommand{\setRow}{\rowcolor{gray!10}}
\newcommand{\mapRow}{} %
\newcommand{\graphRow}{\rowcolor{gray!10}}
\newcommand{\sequenceRow}{} %

\begin{table}[tbp]
    \begin{center}
    \renewcommand{\arraystretch}{.7}
    \small
    \begin{tabularx}{\columnwidth}{@{\extracolsep{0pt}}
					>{\columncolor{gray!10}[0pt][\tabcolsep]\raggedright\arraybackslash}m{4.65cm} 
    					>{\columncolor{gray!10}\centering\arraybackslash}m{0.5cm}
    					>{\columncolor{gray!10}\centering\arraybackslash}m{1cm}
					>{\columncolor{gray!10}\centering\arraybackslash}m{1cm}
					>{\columncolor{gray!10}\raggedleft\arraybackslash}m{1cm}
					>{\columncolor{gray!10}\raggedleft\arraybackslash}m{3.87cm}@{}}
   	 \toprule
	  \rowcolor{white}
        \textbf{CRDT} & \textbf{\!\!\!\!Type} & \textbf{LoC} & \textbf{\!Correct} & \textbf{Time} & \textbf{Source}\\
         \toprule
	Counter & \opBased & 17 & \cmark & \num{3.2} s & \cite{shapiro2011comprehensive} \\ %
        Grow-Only Counter & \stateBased & 33 & \cmark & \num{4.3} s & \cite{shapiro2011comprehensive}\\ %
        Positive-Negative Counter & \stateBased & 15 & \cmark & \num{5.9} s & \cite{shapiro2011comprehensive}\\ %
        \mbox{Dynamic Positive-Negative Counter} & \stateBased & 41 & \cmark & \num{7.1} s & \textcircled{a} \qquad \quad \cite{akkaDynamicPNCounter}\\ \midrule
         \rowcolor{white}
        Enable-Wins Flag & \pure & 18 & \cmark & \num{4.0} s & \cite{Baquero17}\\ %
         \rowcolor{white}
        Disable-Wins Flag & \pure & 20 & \cmark & \num{3.9} s & \cite{Baquero17}\\ \midrule
        Multi-Value Register & \stateBased & 63 & \cmark & \num{8.8} s & \cite{shapiro2011comprehensive}\\ %
        Multi-Value Register & \pure & 18 & \cmark & \num{4.1} s  & \cite{Baquero17}\\ %
        Last-Writer-Wins Register & \stateBased & 16 & \cmark & \num{5.3} s & \cite{shapiro2011comprehensive}\\
        Last-Writer-Wins Register & \opBased & 38 & \cmark & \num{4.4} s & \cite{shapiro2011comprehensive}\\ \midrule
        \rowcolor{white}
        Grow-Only Set & \stateBased & 10 & \cmark & \num{5.3} s & \cite{shapiro2011comprehensive}\\ %
        \rowcolor{white}
        Two-Phase Set & \opBased & 27 & \cmark & \num{4.4} s & \cite{shapiro2011comprehensive}\\ %
       \rowcolor{white}
        Two-Phase Set & \stateBased & 26 & \xmark & \num{6.3} s & \cite{shapiro2011comprehensive}\\ %
        \rowcolor{white}
        Unique Set & \opBased & 39 & \cmark & \num{4.4} s & \cite{shapiro2011comprehensive}\\ %
        \rowcolor{white}
        Add-Wins Set & \pure & 28 & \cmark & \num{4.3} s & \cite{Baquero17}\\ %
        \rowcolor{white}
        Remove-Wins Set & \pure & 42 & \cmark & \num{4.5} s & \cite{Baquero17}\\ %
        \rowcolor{white}
        Last-Writer-Wins Set & \stateBased & 36 & \cmark & \num{6.6} s & \cite{shapiro2011comprehensive}\\ %
        \rowcolor{white}
        Optimized Last-Writer-Wins Set & \stateBased & 37 & \cmark & 6.5 s & new data type\\ %
        \rowcolor{white}
        Positive-Negative Set & \stateBased & 36 & \cmark & \num{9.6} s & \cite{shapiro2011comprehensive}\\ %
        \rowcolor{white}
        Observed-Removed Set & \opBased & 75 & \cmark & \num{6.2} s & \cite{shapiro2011comprehensive}\\ %
        \rowcolor{white}
        Observed-Removed Set & \stateBased & 34 & \cmark & \num{7.6} s &\cite{shapiroStateBasedORSet}\\ %
        \rowcolor{white}
        Optimized Observed-Removed Set & \stateBased & 78 & \cmark & \num{30.2} s & \cite{bieniusa2012optimized}\\ %
        \rowcolor{white}
        Molli, Weiss, Skaf Set & \opBased & 45 & \cmark & \num{4.7} s & \textcircled{i} \cite{shapiro2011comprehensive}\\ \midrule
        Grow-Only Map & \stateBased & 32 & \cmark & \num{9.1} s & new data type\\ %
        Buggy Map & \opBased & 87 & \xmark & \num{65.2} s & \cite{kleppmannMap}\\
        Corrected Map & \opBased & 101 & \cmark & \num{49.4} s & \cite{kleppmannMap}\\
        \midrule
        \rowcolor{white}
        2P2P Graph & \opBased & 58 & \cmark & \num{7.8} s & \cite{shapiro2011comprehensive}\\ %
        \rowcolor{white}
        2P2P Graph & \stateBased & 41 & \cmark & \num{10.7} s & \textcircled{a} \cite{shapiro2011comprehensive}\\ %
        \rowcolor{white}
        Add-Only Directed Acyclic Graph & \opBased & 42 & \cmark & \num{4.7} s & \cite{shapiro2011comprehensive}\\ %
       \rowcolor{white}
        Add-Only Directed Acyclic Graph & \stateBased & 30 & \cmark & \num{8.7} s & \textcircled{a} \cite{shapiro2011comprehensive}\\ \midrule
        Add-Remove Partial Order & \opBased & 61 & \cmark & \num{10.4} s & \cite{shapiro2011comprehensive}\\ %
        Add-Remove Partial Order & \stateBased & 49 & \cmark & \num{13.2} s & \textcircled{a} \cite{shapiro2011comprehensive} \\ %
        Replicated Growable Array & \opBased & 156 & \clock & / & \cite{shapiro2011comprehensive}\\ %
        Continuous Sequence & \opBased & 108 & \cmark & \num{9.2} s & \textcircled{a} %
        \cite{shapiro2011comprehensive}\\ %
        Continuous Sequence & \stateBased & 53 & \cmark & \num{11.4} s & \textcircled{a} \cite{shapiro2011comprehensive}\\
        \bottomrule
    \end{tabularx}
    \end{center}
\caption{\label{tab:crdt-verif}Verification results for CRDTs implemented and verified in \VeriFx.
        S = state-based, O = op-based, P = pure op-based CRDT. 
        \clock\ = timeout, \textcircled{a} = adaptation of an existing CRDT, \textcircled{i} = incomplete definition. %
        }
\end{table}

\Cref{tab:crdt-verif} summarizes the verification results, including the average verification time and code size of the different CRDTs. %
The Dynamic Positive-Negative Counter CRDT is a variation on the traditional Positive-Negative Counter that supports a dynamic number of replicas and is based on the implementation found in Akka's distributed key-value store~\cite{akkaDynamicPNCounter}.
\VeriFx was able to verify all CRDTs except the Replicated Growable Array (RGA)~\cite{shapiro2011comprehensive} due to the recursive nature of the insertion algorithm.
We found that the Two-Phase Set CRDT (described in~\cref{sec:motivation}) converges but is not functionally correct, that the original Map CRDT proposed by \citet{kleppmannMap} diverges as \VeriFx found the counterexample described in their technical report, and that the Molli, Weiss, Skaf (MWS) Set is incomplete. We now focus on the latter.

\begin{figure*}
\centering
\begin{minipage}[t]{.42\textwidth}
\vspace{-2.8mm}
\begin{algorithm}[H]
\caption{Op-based MWS Set CRDT taken from~\citet{shapiro2011comprehensive}.}\label{spec:MWS-Set}
\scriptsize
\begin{algorithmic}[1]
    \State \speckeyword{payload} set $S = \{(\mathit{element}, \mathit{count}), \ldots\}$
 	    \State \indentt \speckeyword{initial} $E \times \{ 0 \}$
 	\query{\emph{lookup} (element $e$) : boolean $b$}
 	    \State \speckeyword{let} $b = ((e,k) \in S \land k > 0)$ \endQuery
 	\update{\emph{add} (element $e$)}
 	    \source {($e$) : integer $j$}
 	        \If { $\exists (e, k) \in S : k \leq 0$ }
              \State \speckeyword{let} $j = |k| + 1$
            \Else \State \speckeyword{let} $j = 1$
            \EndIf
        \endSource
 	    \down{$e, j$}
 	      \State \speckeyword{let} $k' : (e, k') \in S$
 	      \State $S := S \backslash \{(e, k')\} \cup \{(e, k' + j)\}$
 	    \endDown
 	\endUpdate
 	\update{\emph{remove} (element $e$)}
 	    \source{($e$)}
 	      \State \speckeyword{pre} $\mathit{lookup}(e)$
 	    \endSource
 	    \down{$e$}
 	      \State $S := S \backslash \{(e, \mathbf{k'})\} \cup \{(e, \mathbf{k'} - 1)\}$\label{ln:unbound-k}
 	    \endDown
 	\endUpdate
 \end{algorithmic}
\end{algorithm}

\vspace{-6mm}

\begin{algorithm}[H]
\caption{Remove with $k'$ defined at source.}\label{spec:MWS-Set-rmv-1}
\scriptsize
\begin{algorithmic}[1]
 	\update{\emph{remove} (element $e$)}
 	    \source{($e$) : integer $k'$}
 	      \State \speckeyword{pre} $\mathit{lookup}(e)$
 	      \State \speckeyword{let} $k'$ : $(e, k') \in S$
 	    \endSource
 	    \down{$e, k'$}
 	      \State $S := S \backslash \{(e, k')\} \cup \{(e, k' - 1)\}$
 	    \endDown
 	\endUpdate
 \end{algorithmic}
\end{algorithm}

\vspace{-6mm}

\begin{algorithm}[H]
\caption{Remove with $k'$ defined in downstream.}\label{spec:MWS-Set-rmv-2}
\scriptsize
\begin{algorithmic}[1]
 	\update{\emph{remove} (element $e$)}
 	    \source{($e$)}
 	      \State \speckeyword{pre} $\mathit{lookup}(e)$
 	    \endSource
 	    \down{$e$}
 	      \State \speckeyword{let} $k'$ : $(e, k') \in S$
 	      \State $S := S \backslash \{(e, k')\} \cup \{(e, k' - 1)\}$
 	    \endDown
 	\endUpdate
 \end{algorithmic}
\end{algorithm}

\end{minipage}%
\hfill
\begin{minipage}[t]{.51\textwidth}
\begin{lstlisting}[style=VFxStyle, caption={MWS Set implementation in \VeriFx.}, label=lst:MWS-Set, basicstyle={\scriptsize\ttfamily}]
enum SetOp[V] { Add(e: V) | Remove(e: V) }
enum SetMsg[V] { AddMsg(e: V, dt: Int) | RmvMsg(e: V) }
class MWSSet[V](elements: Map[V, Int]) extends CmRDT[SetOp[V], SetMsg[V], MWSSet[V]] {`\label{ln:mws-set-field}'
  override def enabledSrc(op: SetOp[V]) = op match {
    case Add(_) => true
    case Remove(e) => this.preRemove(e) }
  def prepare(op: SetOp[V]) = op match {
    case Add(e) => this.add(e)
    case Remove(e) => this.remove(e) }
  def effect(msg: SetMsg[V]) = msg match {
    case AddMsg(e, dt) => this.addDownstream(e, dt)
    case RmvMsg(e) => this.removeDownstream(e) }
  def lookup(e: V) = this.elements.getOrElse(e, 0) > 0
  def add(e: V): SetMsg[V] = {
    val count = this.elements.getOrElse(e, 0)
    val dt = if (count <= 0) (count * -1) + 1 else 1
    new AddMsg(e, dt) }
  def addDownstream(e: V, dt: Int): MWSSet[V] = {
    val count = this.elements.getOrElse(e, 0)
    new MWSSet(this.elements.add(e, count + dt)) }
  def preRemove(e: V) = this.lookup(e)
  def remove(e: V): SetMsg[V] = new RmvMsg(e)
  def removeDownstream(e: V): MWSSet[V] = {
    val kPrime = ??? // undefined in Specification 2 `\label{ln:unbound-k-impl}'
    new MWSSet(this.elements.add(e, kPrime - 1)) } }
object MWSSet extends CmRDTProof1[SetOp,SetMsg,MWSSet]`\label{ln:mws-proof}'
\end{lstlisting}

\begin{lstlisting}[style=VFxStyle, caption={Computing $k'$ at the source.}, label=lst:MWS-Set-1, basicstyle={\scriptsize\ttfamily}]
def remove(e: V): Tuple[V, Int] =
  new Tuple(e, this.elements.getOrElse(e, 0))
def removeDown(tup: Tuple[V, Int]): MWSSet[V] = {
  val e = tup.fst; val kPrime = tup.snd
  new MWSSet(this.elements.add(e, kPrime - 1)) }
\end{lstlisting}

\begin{lstlisting}[style=VFxStyle, caption={Computing $k'$ downstream.}, label=lst:MWS-Set-2, basicstyle={\scriptsize\ttfamily}]
def remove(e: V): V = e
def removeDown(e: V): MWSSet[V] = {
  val kPrime = this.elements.getOrElse(e, 0)
  new MWSSet(this.elements.add(e, kPrime - 1)) }
\end{lstlisting}

\end{minipage}
\vspace{-4mm}
\end{figure*}

\Cref{spec:MWS-Set} describes the MWS Set, which associates a count to every element.
An element is considered in the set if its count is strictly positive.
\texttt{remove} decreases the element's count, while \texttt{add} increments the count by the amount that is needed to make it positive (or by 1 if it is already positive).
\Cref{lst:MWS-Set} shows the implementation of the MWS Set in \VeriFx as a polymorphic class that extends the \texttt{CmRDT} trait (\cf \cref{sec:CmRDT-lib}).
The type arguments passed to \texttt{CmRDT} correspond to the supported operations (\texttt{SetOp}s), the messages that are exchanged (\texttt{SetMsg}s), and the CRDT type itself (\texttt{MWSSet}).
The \texttt{SetOp} enumeration defines two types of operations: \texttt{Add(e)} and \texttt{Remove(e)}.

The \texttt{MWSSet} class has a field, called \texttt{elements}, that maps elements to their count (Line~\ref{ln:mws-set-field}).
Like all op-based CRDTs, the \texttt{MWSSet} implements two phases: \texttt{prepare} and \texttt{effect}.
\texttt{prepare} pattern matches on the operation and delegates it to the corresponding source method which prepares a \texttt{SetMsg} to be broadcast over the network.
The class overrides the \texttt{enabledSrc} method to implement the source precondition on \texttt{remove}, as defined by~\cref{spec:MWS-Set}.
When replicas receive incoming messages, they are processed by the \texttt{effect} method which delegates them to the corresponding downstream method which performs the actual update.
For example, the \texttt{removeDownstream} method processes incoming \texttt{RmvMsg}s by decreasing some count $k'$ by 1.
Unfortunately, $k'$ is undefined in \cref{spec:MWS-Set}.

We believe that $k'$ is either defined by the source replica and 
included in the propagated message (\cref{spec:MWS-Set-rmv-1}),
or, $k'$ is defined as the element's count at the downstream replica (\cref{spec:MWS-Set-rmv-2}).
We implemented both possibilities in \VeriFx (\cref{lst:MWS-Set-1,lst:MWS-Set-2}) but it is unclear which one, if any, is correct.
To find out, the companion object of the \texttt{MWSSet} class (\cf Line~\ref{ln:mws-proof} in \cref{lst:MWS-Set}) extends the \texttt{CmRDTProof1} trait (\cf \cref{sec:CmRDT-lib}), passing along three type arguments: the type of operations \texttt{SetOp}, the type of messages being exchanged \texttt{SetMsg}, and the CRDT type constructor \texttt{MWSSet}. %
The object extends \texttt{CmRDTProof\textbf{1}} as the \texttt{MWSSet} class is polymorphic and expects one type argument.
When executing the proof inherited by the companion object, \VeriFx automatically proves that the possibility implemented by~\cref{lst:MWS-Set-1} is wrong and that the one of~\cref{lst:MWS-Set-2} is correct.
We thus successfully completed the MWS Set implementation thanks to \VeriFx's integrated verification capabilities.

Based on the figures reported in \cref{tab:crdt-verif} we conclude that \VeriFx is suited to verify CRDTs since all implementations were verified in a matter of seconds. %

\subsection{Verifying Operational Transformation}
\label{sec:eval-ot}
We now show that \VeriFx is general enough to verify other distributed abstractions such as Operational Transformation (OT).
We implemented all transformation functions for collaborative text editing described by~\citet{ImineOT} and verified $\mathit{TP_1}$ and $\mathit{TP_2}$ in \VeriFx.

\begin{figure*}
\begin{minipage}[t]{.51\textwidth}
\begin{footnotesize}
\begin{tabularx}{\linewidth}{@{\extracolsep{0pt}}
    					>{\raggedright\arraybackslash}m{2.84cm}
					>{\centering\arraybackslash}m{0.35cm}
					>{\centering\arraybackslash}m{0.35cm}
					>{\centering\arraybackslash}m{0.35cm}
					>{\raggedleft\arraybackslash}m{0.62cm}
					>{\raggedleft\arraybackslash}m{0.45cm}@{}}

\toprule
\multirow[c]{2}{*}{\makecell{\textbf{Transformation} \\ \textbf{Function}}}&
\multirow[c]{2}{*}{\makecell{\textbf{LoC}}} &
\multicolumn{2}{c}{\textbf{Properties}} &
\multicolumn{2}{c}{\textbf{Time}}\\ \cmidrule(lr){3-4} \cmidrule(lr){5-6}
	& & \multicolumn{1}{c}{$\mathit{TP_1}$} & \multicolumn{1}{c}{$\mathit{TP_2}$} & 
	   \multicolumn{1}{c}{$\mathit{TP_1}$} & \multicolumn{1}{c}{$\mathit{TP_2}$} \\
 \toprule
 \citet{OT} & 84 & \xmark & \xmark & \num{115} s & \num{29} s\\
  \midrule 
 \citet{ResselOT} & 78 & \cmark & \xmark & \num{68} s & \num{30} s \\
  \midrule 
 \citet{SunOT} & 68 & \xmark & \xmark & \num{321} s & \num{13} s \\
  \midrule 
 \citet{SuleimanOT} & 85 & \cmark & \clock &  \num{90} s & / \\
  \midrule 
 \citet{ImineOT} & 83 & \cmark & \xmark & \num{61} s & \num{17} s \\
  \midrule
 Register$_{\mbox{v1}}$ \!\cite{imineRegAndStack} & 6 & \xmark & \cmark & 3 s & 3 s \\
  \midrule 
 Register$_{\mbox{v2}}$ \!\cite{imineRegAndStack} & 6 & \cmark & \xmark & 3 s & 3 s \\
  \midrule 
 Register$_{\mbox{v3}}$\! \cite{imineRegAndStack} & 7 & \cmark & \cmark & 3 s & 3 s \\
  \midrule 
 Stack \cite{imineRegAndStack} & 47 & \xmark & \cmark & 5 s & 5 s \\
 \bottomrule 
\end{tabularx}
\end{footnotesize}
\captionof{table}{\label{tab:OT-results} Verification results of OT functions in \VeriFx. A clock \clock\ indicates that the proof timed out. 
}
\end{minipage}
\hfill
\begin{minipage}[t]{.43\textwidth}
\vspace{-3cm}\begin{lstlisting}[style=VFxStyle, caption={Excerpt from the implementation of \citet{ImineOT}'s transformation functions in \VeriFx.}, label=lst:imine-OT-impl, basicstyle={\scriptsize\ttfamily}]TP_1
enum Op { Ins(p: Int, ip: Int, c: Int) | Del(p: Int) | Id() }`\label{ln:op-adt}'
object Imine extends ListOT[Int, Op] {
  def Tii(x: Ins, y: Ins) = {
    val p1 = x.p; val ip1 = x.ip; val c1 = x.c
    val p2 = y.p; val ip2 = y.ip; val c2 = y.c
    if (p1 < p2) x
    else if (p1 > p2) new Ins(p1 + 1, ip1, c1)
    else if (ip1 < ip2) x
    else if (ip1 > ip2) new Ins(p1+1, ip1, c1)
    else if (c1 < c2) x
    else if (c1 > c2) new Ins(p1+1, ip1, c1)
    else new Id() }
  def Tid(x: Ins, y: Del) =
    if (x.p > y.p) new Ins(x.p - 1, x.ip, x.c)
    else x
  def Tdi(x: Del, y: Ins) =
    if (x.p < y.p) x else new Del(x.p + 1)
  def Tdd(x: Del, y: Del) = if (x.p < y.p) x
    else if (x.p > y.p) new Del(x.p - 1)
    else new Id() }
\end{lstlisting}
\end{minipage}
\vspace{-4mm}
\end{figure*}

\Cref{tab:OT-results} summarizes the verification results for each transformation function and includes the average verification time and code size.
The functions proposed by~\citet{OT} and~\citet{SunOT} do not satisfy $\mathit{TP_1}$ nor $\mathit{TP_2}$.
\Citet{ResselOT}'s functions satisfy $\mathit{TP_1}$ but not $\mathit{TP_2}$.
\Citet{SuleimanOT}'s functions satisfy $\mathit{TP_1}$ but the proof for $\mathit{TP_2}$ times out due to the complexity of the transformations\footnote{Suleiman's transformation functions~\cite{SuleimanOT} do not satisfy $\mathit{TP_2}$ according to \citet{osterOT}.}.
These results confirm prior findings by \citet{ImineOT}.
However, \VeriFx found that the transformation functions proposed by \citet{ImineOT} also do not satisfy $\mathit{TP_2}$, which confirms the findings of \citet{liOT} and \citet{osterOT}.
In addition, in a private communication \citet{imineRegAndStack} asked us to verify (unpublished) OT designs for replicated registers and stacks.
Out of the three register designs verified in \VeriFx, only one was correct for both $\mathit{TP_1}$ and $\mathit{TP_2}$.
Regarding the stack design, it guarantees $\mathit{TP_2}$ but not $\mathit{TP_1}$.
\VeriFx provided meaningful counterexamples for each incorrect design.

To exemplify our approach to verifying OT, we now describe the implementation and verification of \citet{ImineOT}'s transformation functions in \VeriFx, which is shown in \Cref{lst:imine-OT-impl}.
The enumeration \texttt{Op} on Line~\ref{ln:op-adt} defines the three supported operations:
\begin{itemize}
\item \texttt{Ins(p, ip, c)} represents the insertion of character \texttt{c}	 at position \texttt{p}. Initially, the character\footnote{We represent characters using integers that correspond to their ASCII code.} was inserted at position \texttt{ip}. Transformations may change \texttt{p} but leave \texttt{ip} untouched.
\item \texttt{Del(p)} represents the deletion of the character at \mbox{position \texttt{p}.}
\item \texttt{Id()} acts as a no-op. This operation is never issued by users directly but operations may be transformed to a no-op.
\end{itemize}
The object \texttt{Imine} extends the \texttt{ListOT} trait 
and implements the four transformation functions (\texttt{Tii}, \texttt{Tid}, \texttt{Tdi}, \texttt{Tdd}) that are required for collaborative text editing (\cf \cref{sec:ot-library}).
The implementation of these transformation functions is a straightforward translation from their description by \citet{ImineOT}.
The resulting object inherits automated proofs for $\mathit{TP_1}$ and $\mathit{TP_2}$. %
When running these proofs, \VeriFx reports that the transformation functions guarantee $\mathit{TP_1}$ but not $\mathit{TP_2}$.
Based on the results shown in \cref{tab:OT-results}, we conclude that \VeriFx is suited to verify other RDT families such as OT. %
Due to the number of cases that have to be considered, the verification times are longer than for CRDTs but are still acceptable for static verification~\cite{movingFastVerif}.

\section{Discussion}\label{sec:limitations}

Our work explores a 
high-level programming language that is powerful enough to implement distributed abstractions (\eg CRDTs and OT) 
and verify them automatically without requiring annotations or programmer intervention of any kind.
\VeriFx shows that automated verification of RDTs based on SMT solving removes the need for abstract specifications and verifies the actual implementation instead.
Our approach enables programmers to implement RDTs, express correctness properties, and verify those properties automatically, all within the \emph{same} language.
This avoids mismatches between the implementation and the verification without requiring expertise in verification. %
We now discuss the limitations of two key features of our approach.

\paragraph{Traits.} For simplicity, \VeriFx currently only supports single inheritance from traits. %
This could, however, be extended to support multiple inheritance.
Traits are not meant for subtyping because subtyping complicates verification as every subtype needs to be verified but these are not necessarily known at compile time.
Hence,  class fields, method parameters, local variables, etc. cannot be of a trait type. %
Programmers can, however, define enumerations as these have a fixed number of constructors, all of which are known at compile time.
Note that traits can define type parameters with upper type bounds.
These type bounds are only used by the type checker to ensure that every extending class or trait is well-typed. %
The compiled SMT program does not contain traits as they are effectively compiled away (\cf \cref{sec:transformations}).
Proofs, classes, and class methods cannot have type bounds on type parameters because the compiler does not know all subtypes.

\paragraph{Functional collections.}
\VeriFx encodes higher order functions on collections (e.g. \texttt{map}, \texttt{filter}, etc.) using arrays, which are treated as function spaces in the combinatory array logic (CAL)~\cite{z3ArrayTheory}.
Hence, anonymous functions (lambdas) merely define an array and arrays are first-class. %
SMT solvers can efficiently reason about \VeriFx's functional collections and their higher order operations because CAL is decidable.
However, some operations are encoded using universal or existential quantifiers which may hamper decidability.
In practice, we were able to verify RDTs involving complex functional operations.
Unfortunately, \VeriFx's collections do not yet provide aggregation methods (\eg \texttt{fold} and \texttt{reduce}) because this is beyond the capabilities of CAL. %
Instead, programmers need to manually aggregate the collection by writing recursive methods that loop over the values of the collection.
While looping over finite collections works, most SMT solvers will not provide inductive proofs for general properties about recursive functions. %

\section{Related Work}

Automated program verification is a vast area of research.
We focus our comparison of related work on verification languages, and approaches for verifying RDTs, invariants in distributed systems, and operational transformation.

\paragraph{Verification languages.}

Verification languages can be classified in three categories: interactive, auto-active, and automated verification languages~\cite{leino-auto-active-verif}.
Interactive verification languages include proof assistants like Coq and Isabelle/HOL in which programmers define theorems and prove them manually using proof tactics. Although some automation tactics exist, proving complex theorems requires considerable manual proof efforts. Similarly, programmers in Liquid Haskell~\cite{liquidHaskell} provide proofs using plain Haskell functions. Some proofs can be assisted or discharged by the underlying SMT solver.
In contrast, proofs in \VeriFx are fully automated.
Auto-active verification languages like Dafny~\cite{dafny} and Spec\#~\cite{specSharp} verify programs for runtime errors and user-defined invariants based on annotations provided by the programmer (\eg preconditions, postconditions, loop invariants, etc.).
Intermediate verification languages (IVLs) like Boogie~\cite{boogie} and Why3~\cite{why3} automate the proof task by generating verification conditions (VCs) from source code and discharging them using one or more SMT solvers.
IVLs are not meant to be used by programmers directly.
Instead, programs written in some verification language (e.g. Dafny, Spec\#, etc.) are translated to an IVL to verify the VCs.
While the aforementioned approaches aim to be general such that they can be used to prove any property of a program,
\VeriFx was designed to be a high-level programming language capable of verifying RDTs \emph{fully automatically}. %

\paragraph{Verifying SEC for RDTs.}
Much work has focused on verification techniques for RDTs. %
\Citet{burckhardtFormalCRDTFramework} propose a formal framework that enables the specification and verification of RDTs.
\Citet{specificationOfCollabTextEditing} use a variation on this framework to provide precise specifications of replicated lists - which form the basis of collaborative text editing - and prove the correctness of an existing text editing protocol.
\Citet{IsabelleCRDTs} and~\citet{zellerVerificationCRDTs} propose formal frameworks in the Isabelle/HOL theorem prover to mechanically verify SEC for CRDT implementations.
Unfortunately, the aforementioned verification techniques require significant efforts since they are not automated.

\Citet{verificationCRDTsLiquidHaskell} extend Liquid Haskell~\cite{liquidHaskell} with typeclass refinements and use them to prove SEC for several CRDT implementations. While simple proofs can be discharged automatically by the underlying SMT solver, advanced CRDTs also require significant proof efforts (as discussed in \cref{sec:motivation}).

\Citet{abstractionCRDT} propose a new correctness criterion for CRDTs that extends SEC with functional correctness and enables manual verification of CRDT implementations and client programs using them. %
They mainly focus on functional correctness and provide paper proofs.
In contrast, \VeriFx enables \emph{automated} verification of CRDT implementations.

\Citet{ra-linearizability} propose replication-aware linearizability, a criterion that enables sequential reasoning to prove the correctness of CRDT implementations.
The authors manually encoded the CRDTs in the Boogie verification tool to prove correctness.
Those encodings are non-trivial and differ from real-world CRDT implementations.

\Citet{sureshAutomatedVerificationCRDTs} developed a proof rule that is parametrized by the consistency model and automatically checks convergence for CRDTs. %
Unfortunately, their framework introduces imprecisions and may reject correct CRDTs.
Moreover, their framework requires a first-order logic specification of the CRDT. %
The resulting proofs thus verify the specification instead of a concrete implementation.
In contrast, \VeriFx can verify high-level CRDT implementations directly.

Finally, \citet{ECforCRDTs} introduce a notion of validity for RDTs and manually prove it for some CRDTs. We do not consider validity in this work.

\paragraph{Verifying invariants in distributed systems.}

Reasoning about program invariants and maintaining them under weak consistency is challenging.
Invariant confluence~\cite{Bailis14Coordination} is a correctness criterion for coordination avoidance;
invariant confluent operations maintain application invariants, even without coordination.
\Citet{whittaker2018interactive} devise a decision procedure for invariant confluence that can be checked automatically by their interactive system.

Some work has focused on verifying program invariants for existing RDTs \cite{soteria,verifHighlyAvailableProgs,CISE,IPA}.
Soteria~\cite{soteria} verifies program invariants for state-based replicated objects.
Repliss~\cite{verifHighlyAvailableProgs} verifies program invariants for applications that are built on top of their CRDT library.
CISE~\cite{CISE} proposes a proof rule to check that a particular choice of consistency for the operations preserves the application invariants.
IPA~\cite{IPA} detects invariant-breaking operations and proposes modifications to the operations in order to preserve the invariants.
Unfortunately, these approaches assume that the underlying RDT is correct. 
\VeriFx enables programmers to verify that this is the case. %
In this paper, we did not consider application invariants and leave them as future work.

Other approaches~\cite{Sieve, Hamsaz, Hampa, RedBlue, PoR, zhao2018observable, zhao2020replicated, ECROs, mixt} feature a hybrid consistency model that
decides on an appropriate consistency level for operations based on a static analysis.
In this work, we did not consider mixed-consistency RDTs.

\paragraph{Verifying operational transformation functions.}

\Citet{OT} first proposed an algorithm for operational transformation together with a set of transformation functions.
Several works~\cite{SuleimanOTProof,SunOT} showed that integration algorithms like adOPTed~\cite{ResselOT}, SOCT2~\cite{SuleimanOTProof}, and GOTO~\cite{SunOT2} guarantee convergence iff the transformation functions satisfy the $\mathit{TP_1}$ and $\mathit{TP_2}$ properties.
Unfortunately, \citet{OT}'s transformation functions do not satisfy these properties~\cite{SunOT,ResselOT,SuleimanOTProof}.
Over the years, several transformation functions have been proposed~\cite{ResselOT,SunOT,SuleimanOT}.
\Citet{ImineOT} used SPIKE, an automated theorem prover, to verify the correctness of these transformation functions and found counterexamples for all of them, except for \citet{SuleimanOT}'s transformation functions. %
As shown in \Cref{sec:eval-ot}, we were able to reproduce their findings using \VeriFx and generate similar counterexamples.
\Citet{ImineOT} also proposed a simpler set of transformation functions which later was found to also violate $\mathit{TP_2}$~\cite{liOT, osterOT}.
\VeriFx also found this counterexample.

\section{Conclusion}

Replicated data types (RDTs) are widespread among highly available distributed systems but verifying them remains complex, even for experts. %
Automated verification efforts~\cite{verificationCRDTsLiquidHaskell, sureshAutomatedVerificationCRDTs} have been proposed but these cannot yet produce complete correctness proofs from high-level implementations.

To address this issue, we propose \VeriFx, a functional object-oriented programming language
that features a novel proof construct to express correctness properties that are verified automatically.
We leverage the proof construct to build libraries for implementing and verifying two well-known families of RDTs: CRDTs and OT.
Programmers can also implement custom libraries to verify other approaches.
Verified RDT implementations can be transpiled to mainstream languages, \eg Scala or JavaScript.
\VeriFx's modular architecture allows programmers to add support for other languages. %

This work accounts for the first extensive portfolio of verified RDTs including 35 CRDTs and 9 OT designs.
All were verified automatically in a matter of seconds or minutes and with minimal effort. For example, our implementation of an Observed-Removed Set CRDT required only 75 LoC and does not involve verification-specific code. %
In contrast, related work requires significant programmer intervention to verify similar designs, \eg \citet{IsabelleCRDTs} needed 20 auxiliary lemmas to verify the Observed-Removed Set.
\VeriFx allows programmers to implement and automatically verify RDTs within the \emph{same} language, thereby, enabling the adoption of RDTs by the masses. %
\begin{acks}
\noindent 
We would like to thank Abdessamad Imine for the clarifications about transformation functions and his feedback regarding the verification of those transformation functions.
We would also like to thank Christophe Scholliers, Dominique Devriese, Nuno Pregui\c{c}a, and Carlos Baquero for the early feedback and fruitful discussions.
Kevin De Porre was funded by the Research Foundation - Flanders. Project number 1S98519N.
Carla Ferreira was partly funded by EU Horizon Europe under Grant Agreement no. 101093006
(TaRDIS), and FCT-Portugal under grants UIDB/04516/2020 and PTDC/CCI-INF/32081/2017.
\end{acks}

%

\clearpage
\appendix

\section{\VeriFx's Type System}\label{app:type-system}

We now present \VeriFx's type system.
An environment $\tenv$ is a partial and finite mapping from variables to types.
A type environment $\typeEnv$ is a finite set of type variables.
VeriFx's type system consists of a judgment for type wellformdness $\typeWF{\type}$ which says that type $\type$ is well-formed in context $\typeEnv$, and a judgment for typing $\typing{\tenv}{\expr}{\type}$ which says that in context $\typeEnv$ and environment $\tenv$, the expression $\expr$ is of type $\type$.
We abbreviate $\typeWF{\type_1}, \, \ldots, \, \typeWF{\type_n}$ to $\typeWF{\types}$, and
$\typing{\tenv}{\expr_1}{\type_1}, \, \ldots, \, \typing{\tenv}{\expr_n}{\type_n}$ to $\typing{\tenv}{\zeroOrMore{\expr}}{\types}$.

Below we define well-formed types:

$$
\small{
\infer[\textit{WF-String}]{\typeWF{\stringv}}{}
\qquad
\infer[\textit{WF-Bool}]{\typeWF{\boolv}}{}
\qquad
\infer[\textit{WF-Int}]{\typeWF{\intv}}{}
}
$$

$$
\small{
\infer[\textit{WF-Class}]
  { \typeWF{\customType{\className}{\types}} }
  { \typeWF{\types} \\ 
    \classv{\className}{\typeParams}{\ldots}{\ldots} \\
    or \; \; \classvExtends{\className}{\typeParams}{\ldots}{\customType{\traitName}{\ldots}}{\ldots}
  }
}
$$

$$
\small{
\infer[\textit{WF-Trait}]
  { \typeWF{\customType{\traitName}{\types}} }
  { \typeWF{\types} \qquad \types <: \zeroOrMore{\typeP} \\ 
    \trait{\traitName}{\zeroOrMore{\typeParam <: \typeP}}{\ldots} \\
    or \; \; 
    \traitExtends{\traitName}{\zeroOrMore{\typeParam <: \typeP}}{\customType{\traitName}{\ldots}}{\ldots} }
}
$$

$$
\small{
\infer[\textit{WF-Enum}]
  { \typeWF{\customType{\enumName}{\types}} }
  { \typeWF{\types} \\ \adt{\enumName}{\typeParams}{\ldots} }
\qquad
\infer[\textit{WF-TVar}]{\typeWF{\typeParam}}{\typeParam \in \typeEnv}
}
$$

Primitive types are always well-formed.
A type variable $\typeParam$ is valid if it is in scope: $\typeParam \in \typeEnv$, \ie the surrounding method or class defined the type parameter.
Class types and enumeration types are valid if a corresponding class or enumeration definition exists and all type arguments are well-formed.

We now define a few auxiliary definitions which are needed for the typing rules.
The $\mathit{fields}$ function takes a class type and returns its fields and their types:
$$
\small{
\infer[\textsc{F-class}]
{
  \mathit{fields}(\customType{C}{\zeroOrMore{P}}) = [\zeroOrMore{P}/\zeroOrMore{X}] \; \zeroOrMore{v} : \zeroOrMore{\typ}
}
{
  \classv{C}{\zeroOrMore{X}}{\zeroOrMore{v} : \zeroOrMore{\typ}}{\zeroOrMore{M}} \; \; or \; \;
  \classvExtends{\className}{\typeParams}{\zeroOrMore{v} : \zeroOrMore{\typ}}{\customType{\traitName}{\zeroOrMore{\typeQ}}}{\methods} 
}
}
$$

The $\mathit{ftypes}$ function takes an enumeration type and the name of one of its constructors and returns the type of the fields of that constructor.
$$
\small{
\infer[\textsc{FT-enum}]
{
  \mathit{ftypes}(\customType{\enumName}{\zeroOrMore{P}}, \ctorName) = [\zeroOrMore{P}/\typeParams] \; \zeroOrMore{\typ}
}
{
  \adt
    {\enumName}
    {\typeParams}
    {  \ctorName ( \zeroOrMore{v} : \zeroOrMore{\typ} ), \, \ldots }
}
}
$$

The $\mathit{mtype}$ function takes the name of a method and the type of a class, and returns the actual type signature of the method. If the method is not found in the class (\rname{MT-class-rec} rule) it is looked up in the hierarchy of super traits by the \rname{MT-trait} rules. For polymorphic methods, the returned type signature is polymorphic:
$$
\small{
\infer[\textsc{MT-class}]
{
  \mathit{mtype}(m, \customType{C}{\zeroOrMore{P}}) = 
    [\zeroOrMore{P}/\zeroOrMore{X}] \;
    ( \gFunTypeV{\zeroOrMore{Y}}{\zeroOrMore{\typ}}{\typ} )
}
{
  \classv{\className}{\zeroOrMore{X}}{ \ldots }{\zeroOrMore{M}} \; \; or \; \;
  \classvExtends{\className}{\zeroOrMore{X}}{ \ldots }{\customType{\traitName}{\zeroOrMore{\typeQ}}}{\zeroOrMore{M}}
  \\
  \methodv
    {m}
    { \zeroOrMore{Y} }
    { \zeroOrMore{\var} : \zeroOrMore{\typ} }
    { \typ }
    { \expr }
  \in \zeroOrMore{M}
}
}
$$

$$
\small{
\infer[\textsc{MT-class-rec}]
{
  \mathit{mtype}(m, \customType{C}{\zeroOrMore{P}}) = 
    \mathit{mtype}(m, \customType{\traitName}{\zeroOrMore{\typeQ}})
}
{
  \classvExtends{\className}{\zeroOrMore{X}}{ \ldots }{\customType{\traitName}{\zeroOrMore{\typeQ}}}{\zeroOrMore{M}}
  \\
  \methodv
    {m}
    { \zeroOrMore{Y} }
    { \zeroOrMore{\var} : \zeroOrMore{\typ} }
    { \typ }
    { \expr }
  \notin \zeroOrMore{M}
}
}
$$

$$
\small{
\infer[\textsc{MT-trait}]
{
  \mathit{mtype}(m, \customType{\traitName}{\zeroOrMore{\typeP}}) = 
    [\zeroOrMore{P}/\zeroOrMore{X}] \;
    ( \gFunTypeV{\zeroOrMore{Y}}{\zeroOrMore{\typ}}{\typ} )
}
{
  \trait{\traitName}{\zeroOrMore{\typeParam <: \type'}}{\methods} \; \; or \; \;
  \traitExtends{\traitName}{\zeroOrMore{\typeParam <: \type'}}{\customType{\traitName'}{\ldots}}{\methods}
  \\
  \methodv
    {m}
    { \zeroOrMore{Y} }
    { \zeroOrMore{\var} : \zeroOrMore{\typ} }
    { \typ }
    { \expr }
  \in \zeroOrMore{M}
}
}
$$

$$
\small{
\infer[\textsc{MT-trait-rec}]
{
  \mathit{mtype}(m, \customType{\traitName}{\zeroOrMore{\typeP}}) = 
    \mathit{mtype}(m, \customType{\traitName'}{\zeroOrMore{\typeP}})
}
{
  \trait{\traitName}{\zeroOrMore{\typeParam <: \type'}}{\methods} \; \; or \; \;
  \traitExtends{\traitName}{\zeroOrMore{\typeParam <: \type'}}{\customType{\traitName'}{\zeroOrMore{\typeP}}}{\methods}
  \\
  \methodv
    {m}
    { \zeroOrMore{Y} }
    { \zeroOrMore{\var} : \zeroOrMore{\typ} }
    { \typ }
    { \expr }
  \notin \zeroOrMore{M}
}
}
$$

Similarly, we assume that there are functions $valNames(\customType{\traitName}{\zeroOrMore{\typeP}})$ and $declaredMethods(\customType{\traitName}{\zeroOrMore{\typeP}})$ that return all fields, respectively all methods, declared by a trait (and its super traits).
The $\mathit{ctors}$ function takes an enumeration type and returns the names of its constructors.

$$
\small{
\infer[\textsc{C-enum}]
{
  \mathit{ctors}(\customType{\enumName}{\zeroOrMore{P}}) = \oneOrMore{\ctorName}
}
{
  \adt
    { \enumName }
    { \typeParams }
    { \oneOrMore{ \constructor{\ctorName}{ \zeroOrMore{\var} : \zeroOrMore{\typ} }} } 
}
}
$$

\Cref{fig:all-exp-typing-rules} shows the typing rules for expressions.
Most rules are a simplification of Featherweight Generic Java~\cite{fjToplas} without subtyping.
Quantified formulas are boolean expressions if their body also types to a boolean expression in the environment that is extended with the quantified variables (\rname{T-uni} and \rname{T-exi} rules).
Logical implication is a well-typed boolean expression if both the antecedent and the consequent are boolean expressions (\rname{T-impl} rule).

Classes are well-formed if the types of the fields are well-formed and all its methods are well-formed (\rname{T-class1} rule).
If the class extends a trait, it must also implement all fields and methods declared by the hierarchy of super traits (\rname{T-class2} rule). The typing rules for trait definitions and object definitions can be defined similarly.

When instantiating an enumeration through one of its constructors $\new{\customType{\ctorName}{\zeroOrMore{\typeP}}}{\exprs}$, the provided arguments $\exprs$ need to match the types of the constructors' fields, and the result effectively is an object of the enumeration type $\customType{\enumName}{\zeroOrMore{\typeP}}$.

Programmers can pattern match on enumerations but the cases must be exhaustive, \ie every constructor must be matched by at least one case. If all cases are of type $\type$, then the resulting pattern match expression is also of type $\type$.

Finally, the body of a proof must be a well-typed boolean expression.

\begin{figure*}
\centering
$$
\scriptsize{\begin{array}{c}
\infer[\textsc{T-num}]{\typing{\tenv}{\vfx{num}}{\intv}}{}
\quad
\infer[\textsc{T-str}]{\typing{\tenv}{\vfx{str}}{\stringv}}{}
\quad
\infer[\textsc{T-true}]{\typing{\tenv}{\vfx{true}}{\boolv}}{}
\quad
\infer[\textsc{T-false}]{\typing{\tenv}{\vfx{false}}{\boolv}}{}

\\\\[-3pt]

\infer[\textsc{T-var}]
{ 
  \typing
    {\tenv}
    {\var}
    {\tenv(\var)}
}
{
  \var \in \dom{\tenv}
}

\quad

\infer[\textsc{T-neg}]
{
  \typing
    {\tenv}
    { ! \expr }
    {\boolv}
}
{
  \typing
    {\tenv}
    { \expr }
    {\boolv}
}

\quad

\infer[\textsc{T-op1}]
{
  \typing
    {\tenv}
    {\arithmeticOperationVfx{\expr_1}{\expr_2}}
    {\intv}
}
{
  \typing{\tenv}{\expr_1}{\intv}
  \quad
  \typing{\tenv}{\expr_2}{\intv}
}

\\\\[-3pt]

\infer[\textsc{T-op2}]
{
  \typing
    {\tenv}
    {\booleanOperationVfx{\expr_1}{\expr_2}}
    {\boolv}
}
{
  \typing{\tenv}{\expr_1}{\boolv}
  \quad
  \typing{\tenv}{\expr_2}{\boolv}
}

\quad

\infer[\textsc{T-if}]
{
  \typing
    {\tenv}
    { \ifv{\expr_1}{\expr_2}{\expr_3} }
    {\typ}
}
{
  \typing{\tenv}{\expr_1}{\boolv}
  \\
  \typing{\tenv}{\expr_2}{\typ}
  \quad
  \typing{\tenv}{\expr_3}{\typ}
}

\quad

\infer[\textsc{T-val}]
{
  \typing
    {\tenv}
    { \varvv{\var}{\typ_1}{\expr_1}{\expr_2} }
    {\typ_2}
}
{
  \typeWF{\typ_1}\\
  \typing{\tenv}{\expr_1}{\typ_1}
  \quad
  \typing{\tenv, \var : \typ_1}{\expr_2}{\typ_2}
}

\\\\[-3pt]

\infer[\textsc{T-abstraction}]
{
  \typing
    {\tenv}
    { \lambdav{ \oneOrMore{\var} : \oneOrMore{\typ} }{\expr} }
    { \funTypeV{\oneOrMore{\typ}}{\typ} }
}
{
  \typeWF{\oneOrMore{\typ}}\\
  \typing
    {\tenv, \oneOrMore{\var} : \oneOrMore{\typ}}
    {\expr}
    {\typ}
}

\quad

\infer[\textsc{T-call}]
{
  \typing
    {\tenv}
    { \callv{\expr_1}{\oneOrMore{\expr_2}} }
    { \typ }
}
{
  \typing
    {\tenv}
    {\expr_1}
    { \funTypeV{\oneOrMore{\typ}}{\typ} }
  \\
  \typing
    {\tenv}
    {\oneOrMore{\expr_2}}
    {\oneOrMore{\typ}}
}

\quad

\infer[\textsc{T-new-class}]
{
  \typing
    {\tenv}
    { \new{\customType{\className}{\zeroOrMore{\typeP}}}{\exprs} }
    { \customType{\className}{\zeroOrMore{\typeP}} }
}
{
  \mathit{fields}(\customType{\className}{\zeroOrMore{\typeP}}) = \zeroOrMore{\field} : \zeroOrMore{\typ}
  \\
  \typeWF{\customType{\className}{\zeroOrMore{\typeP}}}
  \quad
  \typing
    {\tenv}
    {\zeroOrMore{\expr}}
    { \zeroOrMore{\typ} }
}

\\\\[-3pt]

\infer[\textsc{T-field}]
{
  \typing
    {\tenv}
    { \accessv{\expr}{\field_i} }
    { \typ_i }
}
{
  \typing
    {\tenv}
    {\expr}
    { \typ_o }
  \quad
  \mathit{fields}(\typ_o) = \zeroOrMore{\field} : \zeroOrMore{\typ}
}

\quad

\infer[\textsc{T-invoke}]
{
  \typing
    {\tenv}
    { \invoke{\expr_o}{m}{\zeroOrMore{\typeP}}{\zeroOrMore{\expr}} }
    { [\zeroOrMore{\typeP}/\typeParams] \typ }
}
{
  \typing
    {\tenv}
    {\expr_o}
    { \typ_o } \quad
  \typeWF{\zeroOrMore{\typeP}}
  \\
  \mathit{mtype}(m, \typ_o) = \gFunTypeV{\typeParams}{\zeroOrMore{\typ}}{\typ}
  \\
  \typing
    {\tenv}
    {\zeroOrMore{\expr}}
    { [\zeroOrMore{\typeP}/\typeParams] \zeroOrMore{\typ} }
}

\quad

\infer[\textsc{T-uni}]
{
  \typing
    {\tenv}
    { \forallv{( \oneOrMore{\var} : \oneOrMore{\typ} )} {\expr} }
    {\boolv}
}
{
  \typeWF{\oneOrMore{\typ}}\quad 
  \typing{\tenv, \oneOrMore{\var} : \oneOrMore{\typ}}{\expr}{\boolv}
}

\\\\[-3pt]

\infer[\textsc{T-exi}]
{
  \typing
    {\tenv}
    { \existsv{( \oneOrMore{\var} : \oneOrMore{\typ} )} {\expr} }
    {\boolv}
}
{
  \typeWF{\oneOrMore{\typ}}\quad
  \typing{\tenv, \oneOrMore{\var} : \oneOrMore{\typ}}{\expr}{\boolv}
}

\quad

\infer[\textsc{T-impl}]
{
  \typing
    {\tenv}
    { \impliesv{\expr_1}{\expr_2} }
    {\boolv}
}
{
  \typing{\tenv}{\expr_1}{\boolv}
  \quad
  \typing{\tenv}{\expr_2}{\boolv}
}

\quad

\infer[\textsc{T-new-enum}]
{
  \typing
    {\tenv}
    { \new{\customType{\ctorName}{\zeroOrMore{\typeP}}}{\exprs} }
    { \customType{\enumName}{\zeroOrMore{\typeP}} }
}
{
  \mathit{ctors}( \customType{\enumName}{\zeroOrMore{\typeP}} ) = \oneOrMore{\ctorName} \quad
  \ctorName \in \oneOrMore{\ctorName}\\
  \mathit{ftypes}(\customType{\enumName}{\zeroOrMore{\typeP}}, \ctorName) = \zeroOrMore{\typ}\\
  \typeWF{ \customType{\enumName}{\zeroOrMore{\typeP}} }
  \quad
  \typing
    {\tenv}
    {\zeroOrMore{\expr}}
    { \zeroOrMore{\typ} }
}

\\\\[-3pt]

\infer[\textsc{T-match}]
{
  \typing
    {\tenv}
    { \patternv{\expr_0}{\oneOrMore{c}}  }
    { \typ }
}
{
  \typing
    { \tenv }
    { \expr_0 }
    { \customType{\enumName}{\zeroOrMore{\typeP}} } %
  \\
  ( \mathit{ctors}(\customType{\enumName}{\zeroOrMore{\typeP}}) \setminus \oneOrMore{c} = \emptyset ) \, \vee \,
  ( \casev{\var}{e} \in \oneOrMore{c} ) \, \vee \,
  ( \casev{\_}{e} \in \oneOrMore{c} )
  \\
  \textit{for each } c \in \oneOrMore{c} :
    \typing
      {\tenv}
      {c}
      {\typ}
    \textit{ IN } \patternv{\expr_0}{ \ldots }
}

\\\\[-3pt]

\infer[\textsc{T-ctor-ptn}]
{
  \typing
    { \tenv }
    { \casev{\ctorName(\zeroOrMore{x})}{\expr} }
    { \typ } \textit{ IN } \patternv{\expr_0}{ \ldots }
}
{
  \typing
    { \tenv }
    { \expr_0 }
    { \customType{\enumName}{\zeroOrMore{\typeP}} }
  \\
  \mathit{ftypes}(\customType{\enumName}{\zeroOrMore{\typeP}}, \ctorName) = \zeroOrMore{\typeQ}
  \\
  \typing
    { \tenv, \zeroOrMore{x} : \zeroOrMore{\typeQ} }
    { \expr }
    { \typ }
}

\quad

\infer[\textsc{T-named-ptn}]
{
  \typing
    { \tenv }
    { \casev{\var}{\expr} }
    { \typ } \textit{ IN } \patternv{\expr_0}{ \ldots }
}
{
  \typing
    { \tenv }
    { \expr_0 }
    { \customType{\enumName}{\zeroOrMore{\typeP}} }
  \\
  \typing
    { \tenv, \var : \customType{\enumName}{\zeroOrMore{\typeP}} }
    { \expr }
    { \typ }
}

\\\\[-3pt]

\infer[\textsc{T-wcard-ptn}]
{
  \typing
    { \tenv }
    { \casev{\_}{\expr} }
    { \typ } \textit{ IN } \patternv{\expr_0}{ \ldots }
}
{
  \quad
  \typing
    { \tenv }
    { \expr }
    { \typ }
}
\quad

\infer[\textsc{T-enum}]
{
  \adt
    { \enumName }
    { \typeParams }
    { \oneOrMore {  \constructor{\ctorName}{\zeroOrMore{\field} : \zeroOrMore{\typ}}  } }
    \textit{ OK }
}
{
    \typeEnv = \typeParams \quad
    \typeWF{\zeroOrMore{\typ}}
}

\\\\[-3pt]

\infer[ \textsc{T-method}]
{
  \methodv
    { m }
    { \zeroOrMore{Y} }
    { \zeroOrMore{\var} : \zeroOrMore{\typ} }
    { \typ }
    { \expr }
    \textit{ OK IN  } \customType{C}{\zeroOrMore{X}}
}
{
    \typeEnv = \zeroOrMore{X}, \zeroOrMore{Y} \quad
    \typeWF{\zeroOrMore{\typ}, \typ}
    \\
    \classv
      { C }
      { \typeParams }
      { \ldots }
      { \ldots }
    \; \; or \; \;
    \trait
      { C }
      { \typeParams <: \zeroOrMore{\typeQ} }
      { \ldots}
    \; \; or \; \;
    \traitExtends
      { C }
      { \typeParams <: \zeroOrMore{\typeQ} }
      { \ldots}
      { \ldots}
    \\
    \typing
      { \zeroOrMore{\var} : \zeroOrMore{\typ}, this: \customType{C}{\zeroOrMore{X}} }
      { \expr }
      { \typ }  
}
\\\\[-3pt]
\infer[\textsc{T-proof}]
{
  \proofv
    { \proofName }
    { \typeParams }
    { \expr }
  \textit{ OK }
}
{
    \typeEnv = \typeParams \quad
    \typing
      { \emptyset }
      { \expr }
      { \boolv }
}

\qquad 

\infer[\textsc{T-class1}]
{
  \classv
    { \className }
    { \typeParams }
    { \zeroOrMore{\field} : \zeroOrMore{\typ} }
    { \zeroOrMore{\method} }
    \textit{ OK }
}
{
    \typeEnv = \typeParams \quad
    \typeWF{\zeroOrMore{\typ}}
    \\
    \zeroOrMore{\method} \textit{ OK IN } \customType{\className}{\typeParams}
}

\\\\[-3pt]

\infer[\textsc{T-class2}]
{
  \classvExtends
    { \className }
    { \typeParams }
    { \zeroOrMore{\field} : \zeroOrMore{\typ} }
    { \customType{\traitName}{\zeroOrMore{\typeP}} }
    { \zeroOrMore{\method} }
    \textit{ OK }
}
{
    \typeEnv = \typeParams \qquad
    \typeWF{\zeroOrMore{\typ}} \qquad
    \typeWF{\customType{\traitName}{\zeroOrMore{\typeP}}}
    \\
    \trait{\traitName}{\ldots}{ \valDeclOrMethodDeclOrMethodOrProof } \; \; or \; \;
    \traitExtends{\traitName}{\ldots}{\ldots}{ \valDeclOrMethodDeclOrMethodOrProof }
    \\
    valNames( \customType{\traitName}{\zeroOrMore{\typeP}} ) \subset \zeroOrMore{\field}
    \qquad
    declaredMethods( \customType{\traitName}{\zeroOrMore{\typeP}} ) \subset \methods
    \qquad
    \zeroOrMore{\method} \textit{ OK IN } \customType{\className}{\typeParams}
}

\\\\[-3pt]

\infer[\textsc{T-trait}]
{
  \traitExtends{\traitName}{\zeroOrMore{\typeParam} <: \zeroOrMore{\type}}{ \traitName' \, \langle \, \zeroOrMore{P} \, \rangle }{\zeroOrMore{\valDeclOrMethodDeclOrMethodOrProof}}
  \textit{ OK }
}
{
  \typeEnv = \typeParams \qquad
  \typeWF{\zeroOrMore{\typ}} \qquad
  \typeWF{\customType{\traitName'}{\zeroOrMore{\typeP}}}
  \\
  \trait{\traitName'}{\ldots}{ \ldots } \; \; or \; \;
  \traitExtends{\traitName'}{\ldots}{\ldots}{ \ldots }
  \\
  \valDeclOrMethodDeclOrMethodOrProof = \zeroOrMore{\valDecl} \cup \zeroOrMore{\methodDecl} \cup \methods
  \qquad
  \methods \textit{ OK IN } \customType{\traitName}{\typeParams}
  \\
  valNames( \customType{\traitName'}{\zeroOrMore{\typeP}} ) \subset \zeroOrMore{\valDecl}
  \qquad
  declaredMethods( \customType{\traitName'}{\zeroOrMore{\typeP}} ) \subset ( \zeroOrMore{\methodDecl} \cup \methods )
}

\end{array}}
$$

\caption{Typing \VeriFx expressions.}
\label{fig:all-exp-typing-rules}
\end{figure*}

\clearpage
\section{Core SMT Expressions}\label{app:smt-expressions}

\begin{figure}[b]
$$
\small{
\begin{array}{@{}r@{\,}c@{\,}l@{\quad}l}
\expr & ::= & \smt{num} \, \mid \,  \smt{str} \, \mid \, \smt{true} \, \mid \, \smt{false} & \expl{primitive values} \\
  & \mid & \readsmt{\exprsmt}{\oneOrMore{\exprsmt}} \, \mid \, %
			    \writesmt{\exprsmt}{\oneOrMore{\exprsmt}}{\exprsmt} \, \mid \, %
				\lambdasmt{ \argsmt{\oneOrMore{\param}}{\oneOrMore{\type}} } { \exprsmt } &\\
  & \mid & \varName \, \mid \, \arithmeticOperationSmt{\exprsmt}{\exprsmt} \, \mid \,  \booleanOperationSmt{\exprsmt}{\exprsmt} \, \mid \, \neg \exprsmt &\\
  & \mid & \matchsmt{\exprsmt}{\oneOrMore{\casesmt{\mathit{ptn}} {\exprsmt} }} & \expl{pattern matching}\\
  & \mid & \letsmt{\param}{\exprsmt}{\exprsmt} & \expl{let expression}\\
  & \mid & \ifsmt{\mathit{\exprsmt}}{\exprsmt}{\exprsmt} & \expl{conditional expression}\\
  & \mid & \callsmt{\exprsmt}{\exprsmt} & \expl{function call}\\
  & \mid & \callWithTypessmt{\funName}{\zeroOrMore{\type}}{\exprsmt} &\mbox{\textit{(function call with}} \\
  &  & & \mbox{\textit{\ explicit type arguments)}}\\
  & \mid & \accesssmt{\exprsmt}{\field} & \expl{field access}\\
  & \mid & \forallsmt{ \argsmt{\oneOrMore{\param}} {\oneOrMore{\type}} }{\exprsmt} \, \mid \, \existssmt{ \argsmt{\oneOrMore{\param}} {\oneOrMore{\type}} }{\exprsmt} & \expl{quantified formulas}\\
  & \mid & \impliessmt{\exprsmt}{\exprsmt} & \expl{logical implication}\\
\mathit{ptn} & ::= & \ctorName(\zeroOrMore{\param}) \mid \param & \expl{patterns}
\end{array}
}
$$
\caption{All Core SMT expressions.}
\label{fig:smt-expressions-full}
\end{figure}

We will now discuss the expressions that are supported by Core SMT.
Those expressions are common to most SMT solvers,
except lambdas which, as mentioned before, are described by the preliminary proposal for SMT-LIB v3.0 and are only implemented by some SMT solvers such as Z3~\cite{z3}.

\Cref{fig:smt-expressions-full} provides an overview of all Core SMT expressions.
The simplest expressions are literal values representing integers, strings, and booleans.
Core SMT supports the typical arithmetic operators ($+$, $-$, $*$, $/$) and boolean operators ($\land$, $\lor$, and negation $\neg$) as well as universal and existential quantification, and logical implication.
Immutable variables are defined by let bindings.
Pattern matching is supported but the cases must be exhaustive.
For example, when pattern matching against an algebraic data type every constructor must be handled.
Core SMT supports two types of patterns: constructor patterns $\name(\zeroOrMore{\name})$ that match a specific ADT constructor $\name$ and binds names to its fields $\zeroOrMore{\name}$,
and wildcard patterns that match anything and give it a name $n$.
References $v$ refer to variables that are in scope, \eg function parameters or variables introduced by let binding or pattern matching.
If statements are supported but an else branch is mandatory and both branches must evaluate to the same sort.
Functions can be called and type arguments can be provided explicitly to disambiguate polymorphic functions.
For example, we defined an ADT $\optionType{T}$ with two constructors $\smt{Some}$ and $\smt{None}$.
When calling the $\smt{None}$ constructor we need to explicitly provide a type argument since it cannot be inferred from the call, \eg $\callWithTypessmt{\smt{None}}{\intsmt}{}$.
Finally, fields of an ADT can be accessed by their name. %
Arrays and lambdas were already discussed in \cref{sec:core-smt-syntax}.

\section{Compiler Semantics}

We now discuss the compiler semantics that were not discussed in the main body of the paper.
First, we provide all compilation rules for expressions in \cref{app:compile-expressions}.
Then, we provide all compilation rules for sets and maps in \cref{app:full-set-implementation,app:full-map-implementation} respectively.

\subsection{Compiling Expressions}\label{app:compile-expressions}

\Cref{fig:compiling-expressions} shows the compilation rules for expressions.
The operands of binary operators $\arithmeticOperationVfx{}{}$ are compiled recursively.
A negated expression is compiled to the negation of the compiled expression.
For if statements, the condition and both branches are compiled recursively.
In \VeriFx, $\vfx{this}$ can be used inside the body of a method to refer to the current object.
The reference is compiled to a similar $this$ reference in Core SMT which refers to the $this$ parameter which is always the first parameter of any method (\cf compilation of class methods in \cref{sec:transformations}).
We explained how to compile the remaining expressions in \cref{sec:transformations}.

\begin{figure}[b]
$$
\small{
\begin{array}{lcl}
	\convexp{\var}  & = &  \var \\[0.1em]
 	\convexp{\arithmeticOperationVfx{\expr_1}{\expr_2}} & = & \arithmeticOperationSmt{\convexp{\expr_1}}{\convexp{\expr_2}} \\[0.1em]
	\convexp{!e} & = & \neg \convexp{e} \\[0.1em]
	\convexp{\varvv{\var}{T}{e_1}{e_2}}  & = & \letsmt{\var}{\convexp{e_1}}{ \convexp{e_2}}\\[0.1em]
	\convexp{\ifv{\expr_1}{\expr_2}{\expr_3}}& = & \ifsmt{\convexp{\expr_1}} {\convexp{\expr_2}}{\convexp{\expr_3}} \\[0.1em]
	\convexp{\lambdav{\oneOrMore{\param} : \oneOrMore{\type}} {\expr}}  & = & \lambdasmt{ \oneOrMore{\param} : \oneOrMore{\convtype{\type}} } { \convexp{\expr} } \\[0.1em]
		\convexp{\callv{e_1}{\zeroOrMore{e_2}}} & = & \readsmt{ \convexp{e_1} } { \zeroOrMore{\convexp{e_2}} } \\[0.1em]
	\convexp{\new{\vfx{\setv{T}}}{}} & = & \lambdasmt{ \argsmt{\param}{\convtype{T}} }{\smt{false}} \\[0.1em]
	\convexp{\new{\vfx{\mapv{\type}{\typeP}}}{}} & = & \lambdasmt{\argsmt{\param} {\convtype{\type}}}{\none{\convtype{\typeP}}} \\[0.1em]
	\convexp{\new{\customType{\className}{\zeroOrMore{\type}}}{\exprs}} & = &
	\callWithTypessmt{ \className' } { \zeroOrMore{\convtype{\type}} } { \zeroOrMore{\convexp{\expr}} }\\[0.1em]
	 \multicolumn{3}{l} { \quad \mbox{where } \className' = \concat{\className}{"\_ctor"} } \\[0.1em] %
	\convexp{\new{\customType{\ctorName}{\zeroOrMore{\type}}}{\exprs}} & = &
	\callWithTypessmt{ \ctorName } { \zeroOrMore{\convtype{\type}} } { \zeroOrMore{\convexp{\expr}} }\\[0.1em]
	\convexp{\accessv{\expr}{\field}}  & = & \accesssmt{\convexp{\expr}}{\field} \\[0.1em]
	\convexp{\invoke{\expr_1}{\methodName}{\zeroOrMore{\type}}{\exprs}} & = & %
	  \callWithTypessmt{ \methodName' } { \zeroOrMore{\convtype{\typeP}} , \, \zeroOrMore{\convtype{\type}} } { \convexp{\expr_1}, \, \zeroOrMore{\convexp{\expr}} } \\[0.1em] %
		 \multicolumn{3}{l} { \quad \mbox{where } \mathit{typeof}(e_1) = \className \langle \zeroOrMore{\typeP} \rangle } \\[0.1em]
		 \multicolumn{3}{l} { \quad \mbox{ and } \methodName' = \concatt{\className}{"\_"}{\methodName} \mbox{ and } \zeroOrMore{\typeP} \cap \zeroOrMore{\type} = \emptyset }
\end{array}
}
$$
\caption{Compiling expressions.}
\label{fig:compiling-expressions}
\end{figure}

\Cref{fig:compiling-logical-expressions} shows the compilation rules for logic expressions which in \VeriFx can only occur within the body of proofs.
For quantified formulas the types of the variables $\oneOrMore{\type}$ and the formula $\logicExp$ are compiled.
For logical implications, the antecedent and the consequent are compiled recursively.

\begin{figure}[b]
$$
\small{
\begin{array}{lcl}
\convexp{\forallv { ( \oneOrMore{\var} : \oneOrMore{\type} ) } {\expr}} & = &
		\forallsmt { \argsmt{\oneOrMore{\var}}{\oneOrMore{\convtype{\type}}} } { \convexp{\expr} } \\[0.1em]
\convexp{\existsv { ( \oneOrMore{\var} : \oneOrMore{\type} ) } {\expr}} & = &
		\existssmt { \argsmt{\oneOrMore{\var}}{\oneOrMore{\convtype{\type}}} } { \convexp{\expr} }\\[0.1em]
\convexp{\impliesv{\expr_1}{\expr_2}} & = & \impliessmt { \convexp{\expr_1} } { \convexp{\expr_2} }
\end{array}
}
$$
\caption{Compiling logical expressions.}
\label{fig:compiling-logical-expressions}
\end{figure}

\begin{figure}
$$
\small{
\begin{array}{lcl}
	\convexp{\patternv{\expr}{\oneOrMore{\casev{\patt}{\expr_c}}}}  & = & 
	\matchsmt { \convexp{\expr} } { {\oneOrMore{ \convptn{\casev{\patt}{\expr_c}}}} }\\[0.1em]
	\convptn{\casev{\ctorName (\zeroOrMore{\var})}{\expr}} & = &
	\casesmt { \ctorName ( \zeroOrMore{\var} ) } { \convexp{\expr} }\\[0.1em]
	\convptn{\casev{\var}{\expr}} & = &
	\casesmt { \var } { \convexp{\expr} }\\[0.1em]
	\convptn{\casev{\_}{\expr}} & = &
	\casesmt { \_ } { \convexp{\expr} }
\end{array}
}
$$
\caption{Compiling pattern match expressions.}
\end{figure}

Finally, pattern match expressions are compiled to similar pattern match expressions in Core SMT.
To this end, every pattern is compiled recursively.
Core SMT supports two types of patterns: constructor patterns $\name_1(\zeroOrMore{\name_2})$ that match an algebraic data type constructor $\name_1$ and binds its fields to the provided names $\zeroOrMore{\name_2}$, and wildcard patterns $\name$ that match any value and give it a name $\name$.
Every \VeriFx pattern is compiled to the corresponding Core SMT pattern.
The first pattern, $n_1 ( \zeroOrMore{n_2} )$, matches an ADT constructor $n_1$ and binds its fields to $\zeroOrMore{n_2}$.
It is compiled to an equivalent constructor pattern in Core SMT.
The other two patterns match any expression and are compiled to an equivalent wildcard pattern in Core SMT.

\subsection{Compiling Sets}\label{app:full-set-implementation}

In \cref{sec:transformations} we explained how basic set operations (\texttt{add}, \texttt{remove}, \texttt{contains}) and some advanced operations (\texttt{filter}, \texttt{map}) are compiled to Core SMT.
Now, we explain how the remaining operations on sets are compiled.
\Cref{fig:complete-set-operations} shows the compilation rules for operations over sets.
The union of two sets $e_1$ and $e_2$ is compiled to a lambda which defines an array of elements $v$ of type $\convtype{T}$ containing only elements that are in at least one of the two sets, \ie $\readsmt{\convexp{e_1}}{v} \vee \readsmt{\convexp{e_2}}{v}$.
Similarly, the intersection of two sets $e_1$ and $e_2$ is compiled to a lambda which defines an array containing only elements that are in both sets, \ie $\readsmt{\convexp{e_1}}{v} \wedge \readsmt{\convexp{e_2}}{v}$.
For set difference, the lambda defines an array containing only elements that are in $e_1$ and not in $e_2$. %
A set $e_1$ is a subset of $e_2$ iff all elements from $e_1$ are also in $e_2$.
A set $e$ is non empty if an element $v$ exists that is in the set, \ie $\readsmt{\convexp{e}}{v}$.
A set $e$ is empty if all elements $v$ are not in the set.
A predicate $e_2 : \funTypeV{T}{\boolv}$ holds for all elements of a set $e_1$ if for every element $v$ that is in the set the predicate is true, \ie $\readsmt{\convexp{e_1}}{ v } \implies \readsmt{\convexp{e_p}}{v}$.
A predicate $e_2 : \funTypeV{T}{\boolv}$ holds for at least one element of a set $e_1$ if there exists an element $v$ that is in the set and for which the predicate holds, \ie $\readsmt{\convexp{e_1}}{v} \wedge \readsmt{\convexp{e_p}}{v}$.

\begin{figure*}
\small{
\begin{align*}
	& \convset{e_1.\mathit{add}(e_2)} = 
	\writesmt {\convexp{e_1}} { \convexp{e_2} } { \smt{true} }\\[-0.3em]
	& \convset{e_1.\mathit{remove}(e_2)} =
	\writesmt {\convexp{e_1}} { \convexp{e_2} } { \smt{false} }\\[-0.3em]
	& \convset{e_1.\mathit{contains}(e_2)} =
	\readsmt {\convexp{e_1}} {\convexp{e_2}} \\[-0.3em]
	& \convset{e_1.\mathit{filter(e_2)}} =
		 \lambdasmt{ \argsmt{\varName}{\convtype{T}} } { \readsmt{\convexp{e_1}}{\varName} \wedge \readsmt{\convexp{e_2}}{\varName} }
		\quad \mbox{where } \mathit{typeof}(e_1) =  \mathtt{Set} \langle T \rangle
		 \\[-0.3em]
    &\convset{\expr_1.\mathit{map(\expr_2)}} =
      \lambdasmt{ \argsmt{\paramY}{\convtype{\typeP}} } {  
        \existssmt{\argsmt{\varName}{\convtype{\type}}} {
           \readsmt{\convexp{\expr_1}}{\varName} \wedge \readsmt{\convexp{\expr_2}}{\varName} = \paramY} } \\[-0.3em]
           &
		  \quad \mbox{where } \mathit{typeof}(\expr_1) =  \mathtt{Set} \langle \type \rangle \wedge \mathit{typeof}(\expr_2) =  \type \rightarrow \typeP \\[-0.3em]
	&\convset{e_1.\mathit{union(e_2)}} =
	  \lambdasmt
	    {  \argsmt{\var}{\convtype{T}} }
	    { \readsmt{\convexp{e_1}}{\var} \vee \readsmt{\convexp{e_2}}{\var} }
		\quad \mbox{where } \mathit{typeof}(e_1) =  \mathtt{Set} \langle T \rangle \wedge \mathit{typeof}(e_2) =  \mathtt{Set} \langle T \rangle \\[-0.3em]
	&\convset{e_1.\mathit{intersect(e_2)}} =
	  \lambdasmt
	    {  \argsmt{\var}{\convtype{T}} }
	    { \readsmt{\convexp{e_1}}{\var} \wedge \readsmt{\convexp{e_2}}{\var} } %
		\quad \mbox{where } \mathit{typeof}(e_1) =  \mathtt{Set} \langle T \rangle \wedge \mathit{typeof}(e_2) =  \mathtt{Set} \langle T \rangle \\[-0.3em]
	&\convset{e_1.\mathit{diff(e_2)}} =
	  \lambdasmt
	    {  \argsmt{\var}{\convtype{T}} }
	    { \readsmt{\convexp{e_1}}{\var} \wedge \neg \readsmt{\convexp{e_2}}{\var} } %
	    \quad \mbox{where } \mathit{typeof}(e_1) =  \mathtt{Set} \langle T \rangle \wedge \mathit{typeof}(e_2) =  \mathtt{Set} \langle T \rangle \\[-0.3em]
	&\convset{e_1.\mathit{subsetOf(e_2)}} =
	  \forallsmt
	    { \argsmt{\var}{\convtype{T}} }
	    { \impliessmt
	        {\readsmt{\convexp{e_1}}{\var}}
	        {\readsmt{\convexp{e_2}}{\var}}}
	    \quad \mbox{where } \mathit{typeof}(e_1) =  \mathtt{Set} \langle T \rangle \wedge \mathit{typeof}(e_2) =  \mathtt{Set} \langle T \rangle \\[-0.3em]
	&\convset{e.\mathit{nonEmpty()}} =
	\existssmt { \argsmt{\var}{\convtype{T}} } { \readsmt{\convexp{e}}{\var} } \quad \mbox{ where } \mathit{typeof}(e) =  \mathtt{Set} \langle T \rangle \\[-0.3em]
	&\convset{e.\mathit{isEmpty()}} = 
	\forallsmt { \argsmt{\var}{\convtype{T}} } {\neg \readsmt{\convexp{e}}{\var} } \quad \mbox{ where } \mathit{typeof}(e) =  \mathtt{Set} \langle T \rangle \\[-0.3em]
	&\convset{e_1.\mathit{forall(e_p)}} =
	\forallsmt { \argsmt{\var}{\convtype{T}} } { \readsmt{\convexp{e_1}}{ \var } \implies \readsmt{\convexp{e_p}}{\var} } \\[-0.3em]
		 & \quad \mbox{where } \mathit{typeof}(e_1) =  \mathtt{Set} \langle T \rangle \mbox{ and }
		\mathit{typeof}(e_p) =  \funTypeV{T}{\boolv} \\[-0.3em]
	&\convset{e_1.\mathit{exists(e_p)}} = 
	\existssmt { \argsmt{\var}{\convtype{T}} } { \readsmt{\convexp{e_1}}{\var} \wedge \readsmt{\convexp{e_p}}{\var} } %
		\quad \mbox{where } \mathit{typeof}(e_1) =  \mathtt{Set} \langle T \rangle \mbox{ and }
		\mathit{typeof}(e_p) =  \funTypeV{T}{\boolv}
\end{align*}
}
\caption{Compiling set operations.}
\label{fig:complete-set-operations}
\end{figure*}

\subsection{Compiling Maps}\label{app:full-map-implementation}

\Cref{sec:map-encoding} explained how to encode maps in SMT using arrays and how to efficiently encode the basic map operations.
We now explain how to encode the advanced map operations.
\Cref{fig:advanced-map-ops} defines the SMT encoding for all advanced map operations. %
The \texttt{keys} method on maps returns a set containing only the keys that are present in the map.
Calls to \texttt{keys} on a map $e_m$ of type $\mapv{K}{V}$ are compiled to a lambda which defines a set of keys $k$ of the compiled key type $\convtype{K}$ such that a key is present in the set iff it is present in the compiled map:
$ \readsmt{\convexp{e_m}}{k} \neq \none{\convtype{V}} $.
A predicate $e_p$ of type $\funTypeV{(K, V)}{\boolv}$ holds for all elements of a map $e_m$ of type $\mapv{K}{V}$ iff it holds for every key $k$ that is present in the map and its associated value:
\small{\[ \underbrace{\readsmt{\convexp{e_m}} {k} \neq \none{\convtype{V}}}_\text{$e_m.\mathit{contains}(k)$} \implies \underbrace{\readsmt{\convexp{e_p}} {k, \accesssmt{\readsmt{\convexp{e_m}} {k}} {\mathit{value}} } }_\text{$\callv{e_p}{k, e_m.\mathit{get}(k)}$} \]}
Similarly, the \texttt{values} method returns a set containing all values of the map. To this end, it defines an array containing all values for which at least one key exists that maps to that value.

A predicate $e_p$ of type $\funTypeV{(K, V)}{\boolv}$ holds for at least one element of a map $e_m$ of type $\mapv{K}{V}$ iff there exists a key $k$ with associated value $v$ that is present in the map and for which the predicate holds.
Mapping a function $e_f$ over the key-value pairs of a map $e_m$ is encoded as a lambda that defines an array containing only the keys that are present in the compiled map $\convexp{e_m}$ and whose values are the result of applying $e_f$ on the original value, \ie $\some{ \readsmt{\convexp{e_f}} {k, \accesssmt{\readsmt{\convexp{e_m}} {k}} {\mathit{value}} } }$.
The \texttt{mapValues} method is similar except that it applies the provided function only on the value.
A map $e_m$ can be filtered using a predicate $e_p$ such that the resulting map only contains key-value pairs that fulfill the predicate. Calls to \texttt{filter} are encoded as a lambda that defines an array containing only the key-value pairs that are in the compiled map:
\small{
\begin{align*}
& \ifsmt { \overbrace{\readsmt{\convexp{e_m}} {k} \neq \none{\convtype{V}}}^\text{in original map} \; \wedge \overbrace{\readsmt{\convexp{e_p}} {k, \accesssmt{\readsmt{\convexp{e_m}}{k}} {\mathit{value}} }}^\text{predicate holds} }
 { \\ & \quad \underbrace{\some{ \accesssmt{\readsmt{\convexp{e_m}} { k }} {\mathit{value}} } }_\text{then keep the value} }
 { \, \underbrace{\none{\convtype{V}} }_\text{else not in the map} }
\end{align*}
}
To zip two maps $e_{m_1}$ and $e_{m_2}$ the compiler creates a lambda that defines an array containing only the keys that are present in both maps and the value is a tuple holding the corresponding values from both maps:
\small{\[ \some { \callsmt{Tuple\_ctor} { 
\accesssmt{\readsmt{\convexp{e_{m_1}}} { k }} {\mathit{value}}, \accesssmt{\readsmt{\convexp{e_{m_2}}} { k }} {\mathit{value}} } } \]}
To combine two maps $e_{m_1}$ and $e_{m_2}$ with a function $e_f$ the compiler creates a lambda that defines an array containing all the keys from $e_{m_1}$ and $e_{m_2}$. If a key is present in both maps their values are combined using the provided function $e_f$:
\small{\[ \some { \readsmt{\convexp{e_f}} { \accesssmt{\readsmt{\convexp{e_{m_1}}} { k }} {\mathit{value}}, \accesssmt{\readsmt{\convexp{e_{m_2}}} { k }} {\mathit{value}} } } \]}
If a key-value pair is present in only one of the maps it is also present in the new map.
If a key is not present in $e_{m_1}$ neither in $e_{m_2}$ then it is also not present in the resulting map.

\begin{figure*}
\small
\begin{align*}
  &\convmap{e_m.\mathit{keys}()} =
  \lambdasmt{ \argsmt{\var}{\convtype{K}} } { \readsmt{\convexp{e_m}}{\var} \neq \none{\convtype{V}} }
\quad \mbox{where } \mathit{typeof}(e_m) = \mathtt{Map} \langle K, V \rangle \\[-0.3em]
  &\convmap{e_m.\mathit{values}()} =
  \lambdasmt{ \argsmt{\var}{\convtype{V}} } { \existssmt{\argsmt{k}{\convtype{K}}}{ \readsmt{\convexp{e_m}}{k} = \some{\var} } }
\quad \mbox{where } \mathit{typeof}(e_m) = \mathtt{Map} \langle K, V \rangle \\[-0.3em]
  &\convmap{e_m.\mathit{bijective}()} =
  \forallsmt
    { \argsmt{k_1}{\convtype{K}}, \, \argsmt{k_2}{\convtype{K}} }
    { \\[-0.3em] & \qquad \qquad \qquad \qquad \qquad \quad
      ( k_1 \neq k_2 \, \land \,
      \readsmt{\convexp{e_m}}{k_1} \neq \none{\convtype{V}} \, \land \,
      \readsmt{\convexp{e_m}}{k_2} \neq \none{\convtype{V}} )
      \\[-0.3em] & \qquad \qquad \qquad \qquad \qquad \quad
      \, \implies \,
      \readsmt{\convexp{e_m}}{k_1} \neq \readsmt{\convexp{e_m}}{k_2}
    }\\[-0.3em]
& \quad \mbox{where } \mathit{typeof}(e_m) = \mathtt{Map} \langle K, V \rangle \\[-0.3em]
  &\convmap{e_m.\mathit{forall}(e_p)} =
  \forallsmt { \argsmt{\var}{\convtype{K}} }
{ \readsmt{\convexp{e_m}} {\var} \neq \none{\convtype{V}} \implies \readsmt{\convexp{e_p}} {\var, \accesssmt{\readsmt{\convexp{e_m}} {\var}} {\mathit{value}} } }\\[-0.3em]
  & \quad \mbox{where } \mathit{typeof}(e_m) = \mathtt{Map} \langle K, V \rangle
\mbox{ and } \mathit{typeof}(e_p) = \funTypeV{(K, V)}{\boolv} \\[-0.3em]
  &\convmap{e_m.\mathit{exists}(e_p)} =
  \existssmt { \argsmt{\var}{\convtype{K}} }
{ \readsmt{\convexp{e_m}} {\var} \neq \none{\convtype{V}} \; \wedge \; \readsmt{\convexp{e_p}} {\var, \accesssmt{\readsmt{\convexp{e_m}} {\var}} {\mathit{value}}} }\\[-0.3em]
  & \quad \mbox{where } \mathit{typeof}(e_m) = \mathtt{Map} \langle K, V \rangle
\mbox{ and } \mathit{typeof}(e_p) = (K, V) \rightarrow \vfx{bool} \\[-0.3em]
  &\convmap{e_m.\mathit{map}(e_f)} =
 \lambdasmt{ \argsmt{\var}{\convtype{K}} }{\ 
\ifsmt{\readsmt{\convexp{e_m}} {\var} \neq \none{\convtype{V}} }
    { \\[-0.3em] & \qquad \qquad \qquad \qquad \qquad \quad \some{ \readsmt{\convexp{e_f}} {\var, \accesssmt{\readsmt{\convexp{e_m}} {\var}} {\mathit{value}} } } }
    { \\[-0.3em] &  \qquad \qquad \qquad \qquad \qquad \quad \none{\convtype{W}} } }\\[-0.3em]
    & \quad \mbox{where } \mathit{typeof}(e_m) = \mathtt{Map} \langle K, V \rangle
\mbox{ and } \mathit{typeof}(e_f) = (K, V) \rightarrow W \\[-0.3em]
  &\convmap{e_m.\mathit{mapValues}(e_f)} = 
   \lambdasmt{ \argsmt{\var}{\convtype{K}} } {\
     \ifsmt { \readsmt{\convexp{e_m}}{\var} \neq \none{\convtype{V}} }
    { \\[-0.3em] &  \qquad \qquad \qquad \qquad \qquad \quad \some{\readsmt{\convexp{e_f}} {\accesssmt{\readsmt{\convexp{e_m}}{\var}} {\mathit{value}}} } }
    { \\[-0.3em] &  \qquad \qquad \qquad \qquad \qquad \quad \none{\convtype{W}} }
  }\\[-0.3em]
    & \quad \mbox{where } \mathit{typeof}(e_m) = \mathtt{Map} \langle K, V \rangle
\mbox{ and } \mathit{typeof}(e_f) = V \rightarrow W \\[-0.3em]
  &\convmap{e_m.\mathit{filter}(e_p)} =
  \lambdasmt{ \argsmt{\var}{\convtype{K}} } {\
    \ifsmt { \readsmt{\convexp{e_m}} {\var} \neq \none{\convtype{V}} \; \wedge \; \readsmt{\convexp{e_p}} {\var, \accesssmt{\readsmt{\convexp{e_m}}{\var}} {\mathit{value}} } }
    { \\[-0.3em] & \qquad \qquad \qquad \qquad \qquad \quad \some{ \accesssmt{\readsmt{\convexp{e_m}} { \var }} {\mathit{value}} } }
    { \\[-0.3em] & \qquad \qquad \qquad \qquad \qquad \quad \none{\convtype{V}} }
  }\\[-0.3em]
    & \quad \mbox{where } \mathit{typeof}(e_m) = \mathtt{Map} \langle K, V \rangle
\mbox{ and } \mathit{typeof}(e_p) = (K, V) \rightarrow \vfx{bool} \\[-0.3em]
  &\convmap{e_{m_1}.\mathit{zip}(e_{m_2})} =
  \lambdasmt{ \argsmt{\var}{\convtype{K}} } {\ 
    \ifsmt { \readsmt{\convexp{e_{m_1}}} { \var } \neq \none{\convtype{V}} \; \wedge \; \readsmt{\convexp{e_{m_2}}} { \var } \neq \none{\convtype{W}} }
    { \\[-0.3em] & \qquad \qquad \qquad \qquad \qquad \quad \some { \callsmt{Tuple\_ctor} {
      \accesssmt{\readsmt{\convexp{e_{m_1}}} { \var }} {\mathit{value}}, \accesssmt{\readsmt{\convexp{e_{m_2}}} { \var }} {\mathit{value}} } } }
    { \\[-0.3em] & \qquad \qquad \qquad \qquad \qquad \quad \none{\convtype{\vfx{Tuple} \langle V, W \rangle}} } }\\[-0.3em]
    & \quad \mbox{where } \mathit{typeof}(e_{m_1}) = \mathtt{Map} \langle K, V \rangle
\mbox{ and } \mathit{typeof}(e_{m_2}) = \mathtt{Map} \langle K, W \rangle \\[-0.3em]
  &\convmap{e_{m_1}.\mathit{combine}(e_{m_2}, e_f)} =
   \lambdasmt{ \argsmt{\var}{\convtype{K}}}{
       \ifsmt
      { \readsmt{\convexp{e_{m_1}}} {\var} \neq \none{\convtype{V}} \; \wedge \; \readsmt{\convexp{e_{m_2}}} {\var} \neq \none{\convtype{V}} }
    { \\[-0.3em] & \qquad \qquad \qquad \qquad \qquad \quad \some { \readsmt{\convexp{e_f}} { \accesssmt{\readsmt{\convexp{e_{m_1}}} { \var }} {\mathit{value}}, \accesssmt{\readsmt{\convexp{e_{m_2}}} { \var }} {\mathit{value}} } } }
    {
      \\[-0.3em] & \qquad \qquad \qquad \qquad \qquad \quad
        \ifsmt
        { \readsmt{\convexp{e_{m_1}}} { \var } \neq \none{\convtype{V}} }
      { \\[-0.3em] & \qquad \qquad \qquad \qquad \qquad \quad \quad \readsmt{\convexp{e_{m_1}}} { \var } }
      { \\[-0.3em] & \qquad \qquad \qquad \qquad \qquad \quad \quad
        \ifsmt
        { \readsmt{\convexp{e_{m_2}}} { \var } \neq \none{\convtype{V}} }
        { \\[-0.3em] & \qquad \qquad \qquad \qquad \qquad \quad \quad \quad \readsmt{\convexp{e_{m_2}}} { \var } }
        { \\[-0.3em] & \qquad \qquad \qquad \qquad \qquad \quad \quad \quad \none{\convtype{V}} }
      }
    }} \\[-0.3em]
    & \quad \mbox{where } \mathit{typeof}(e_{m_1}) = \mathtt{Map} \langle K, V \rangle
\mbox{ and } \mathit{typeof}(e_{m_2}) = \mathtt{Map} \langle K, V \rangle
\mbox{ and } \mathit{typeof}(e_f) = (V, V) \rightarrow V\\[-0.3em]
  &\convmap{e_m.\mathit{toSet}()} =
      \lambdasmt
        { \argsmt{\var}{\smt{Tuple} \langle \convtype{K}, \convtype{V} \rangle} }
        { \readsmt{ \convexp{e_m} }{ \accesssmt{\var}{fst} } = \some{\accesssmt{\var}{snd}} }
    \quad \mbox{where } \mathit{typeof}(e_m) = \mathtt{Map} \langle K, V \rangle
\end{align*}
\normalsize
\caption{Compiling advanced map operations.}
\label{fig:advanced-map-ops}
\end{figure*}

\subsection{Compilation Example}\label{app:compilation-example}

\Cref{fig:class-compilation-example} shows a concrete example of a polymorphic set implemented in \VeriFx and its compiled code in Core SMT.
The $MSet$ class defines a type parameter $V$ corresponding to the type of elements it holds.
It also contains one field $set$ of type $\customType{Set}{V}$ and defines a polymorphic method $map$ that takes a function $f: \funTypeV{V}{W}$ and returns a new $MSet$ that results from applying $f$ on every element.
The compiled Core SMT code defines an ADT $MSet$ with one type parameter $V$ and one constructor $MSet\_ctor$.
The constructor defines one field $set$ of sort $\arrayType{V}{\boolsmt}$ which is the compiled sort for sets.
In addition, a polymorphic $MSet\_map$ function is defined which takes two type parameters $V$ and $W$ which correspond to $MSet$'s type parameter and $map$'s type parameter respectively.
The function takes two arguments, the object that receives the call and the function $f$.
The function's body calls the $MSet$ constructor with the result of mapping $f$ over the set.

\begin{figure*}
\centering
\begin{subfigure}[b]{.32\textwidth}
  \centering
\begin{lstlisting}[style=VFxStyle, basicstyle={\scriptsize\ttfamily}, frame=none, numbers=none]
class MSet[V](set: Set[V]) {
  def map[W](f: V => W): MSet[W] =
  	new MSet(this.set.map(f))
}
\end{lstlisting}
  \caption{A polymorphic class in \VeriFx.}
  \label{fig:class-vfx-example}
\end{subfigure}%
\begin{subfigure}[b]{.68\textwidth}
  \centering
  $$
  \small{\begin{array}{l}
  \adtsmt
    {MSet}
    { V }
    { \, \ctorsmt{MSet\_ctor}{ \argsmt{set}{\arrayType{V}{\boolsmt}} } \, } \\ %
  \funDefsmt
  {MSet\_map}
  { V, \, W }
  {  \argsmt{this}{\customType{MSet}{V}}, \, \argsmt{f}{\arrayType{V}{W}} }
  { \customType{MSet}{W}}
  { \\ \quad 
    \callsmt
      {MSet\_ctor}
      { \\ \quad \quad
        \lambdasmt
          { \argsmt{\paramY}{W} }
          { %
            \existssmt{ \argsmt{\param}{V} }{ %
              \readsmt{\accesssmt{this}{set}}{\param} \wedge \readsmt{f}{\param} = \paramY } }

      }
  }
  \end{array}}
  $$
  \caption{Compiled Core SMT code.}
  \label{fig:class-compiled-example}
\end{subfigure}
\caption{Example of a polymorphic class in \VeriFx and the compiled code in Core SMT.}
\label{fig:class-compilation-example}
\vspace{-4mm}
\end{figure*}

\end{document}